\documentclass[fleqn,usenatbib]{mnras}

\usepackage{newtxtext,newtxmath}
\usepackage[T1]{fontenc}
\DeclareRobustCommand{\VAN}[3]{#2}
\let\VANthebibliography\thebibliography
\def\thebibliography{\DeclareRobustCommand{\VAN}[3]{##3}\VANthebibliography}

\usepackage{graphicx}	
\usepackage{amsmath}	

\usepackage{xspace}
\usepackage[dvipsnames]{xcolor}
\usepackage{cancel}
\usepackage{comment}

\usepackage{soul}
\usepackage[normalem]{ulem}

\newcommand{\Lya}{\,\ifmmode{{\mathrm{Ly}\alpha}}\else Ly$\alpha$\fi\xspace}
\newcommand{\HI}{{\text{H\MakeUppercase{\romannumeral 1}}}\xspace}
\newcommand{\cm}{\,\ifmmode{{\mathrm{cm}}}\else cm\fi}

\newcommand{\ergps}{\,{\rm erg}\,{\rm s}\ifmmode{}^{-1}\else${}^{-1}$\fi}
\newcommand{\Mpch}{\,{\rm Mpc}\,\ifmmode h^{-1}\else $h^{-1}$\fi}
\newcommand{\snru}{\,\ifmmode{\mathrm{Myr}^{-1}}\else Myr${}^{-1}$\fi}
\newcommand{\kms}{\,\ifmmode{\mathrm{km}\,\mathrm{s}^{-1}}\else km\,s${}^{-1}$\fi\xspace}

\defcitealias{almadamonter_crossing_2024}{AMG24}

\title[Anisotropic \Lya escape]{Lyman-$\alpha$ Escape through Anisotropic Media}

\author[Almada Monter et al.]{
Silvia Almada Monter$^{1,2}$\thanks{E-mail: almada@mpa-garching.mpg.de}, Max Gronke$^{3,1}$, Seok-Jun Chang$^{1}$
\\
$^{1}$ Max Planck Institute for Astrophysics, 85748 Garching, Germany\\
$^{2}$ Ludwig-Maximilians-Universität München, Geschwister-Scholl-Platz 1, 80539 München, Germany\\
$^{3}$ Astronomisches Rechen-Institut, Zentrum für Astronomie, Universität Heidelberg, Mönchhofstraße 12-14, 69120 Heidelberg, Germany
}

\date{Accepted on Feb 13, 2026}

\pubyear{2026}

\begin{document}
\label{firstpage}
\pagerange{\pageref{firstpage}--\pageref{lastpage}}
\maketitle

\begin{abstract}
The escape of Lyman-$\alpha$ (\Lya) radiation encodes valuable information on the neutral interstellar medium and is often used as a proxy ionizing photon escape. Yet, the theory of \Lya transfer through anisotropic gas distributions remains underdeveloped. We present Monte Carlo radiative transfer simulations of \Lya propagation through porous, inhomogeneous neutral gas, systematically exploring the effects of channel geometry, outflows, dust, and lognormally distributed column densities. 
We find that \Lya photons do not preferentially escape through the lowest-column-density pathways, but instead traverse gas of substantial optical depth, leading to suppressed central flux and the absence of strongly beamed escape.
Subdividing channels has little impact, indicating geometry and covering fraction are more important than porosity.
Channels containing moderate amounts of neutral hydrogen alter escape in characteristic ways, including the appearance of quadruple-peaked spectra, which can be captured by a simple flux–channel relation.
Outflows reshape the spectra by facilitating escape through dense media, redshifting photons, and blending central features, while dust modulates the visibility of small channels by suppressing flux at line center; in both cases, we develop an analytical model that predicts the resulting central fluxes.
Extending to lognormal column density fields, we show \Lya photons probe a broad range of optical depths, producing skewed spectra that can be approximated by weighted sums of homogeneous models. 
Our results have direct implications for using \Lya as a tracer of gas properties and ionizing photon escape; for instance, spectra suggestive of high column densities may nonetheless allow LyC leakage through narrow channels.
\end{abstract}

\begin{keywords}
line: formation -- line:profiles -- radiative transfer -- scattering -- galaxies: high-redshift -- galaxies: ISM 
\end{keywords}


\section{Introduction}
Lyman-alpha (\Lya) is the brightest spectral emission line in the Universe and serves as a crucial observational tool in extragalactic astronomy. It is widely used to study the interstellar and circumgalactic media (ISM and CGM) of galaxies and to detect the most distant galaxies known to date \citep[e.g.,][]{partridge_are_1967,barnes_ly-_2014,dijkstra_lyman_2014,ouchi_observations_2020}. \Lya is a resonant transition, meaning that as \Lya photons propagate through neutral hydrogen (HI)–rich media, they undergo numerous scatterings that change both their spatial paths and their frequencies, imprinting the properties of the surrounding gas onto the emergent signal \citep[see review by][]{dijkstra19}. These radiative transfer effects give rise to the wide observational diversity of \Lya. Spectra can range from narrow, single-peaked emission to complex profiles with multiple peaks, broad wings, strong asymmetries, and large velocity offsets from systemic \citep[e.g.,][]{erb_ly_2014,hayes_spectral_2021,runnholm_lyman_2021}. The same scattering processes also redistribute photons spatially, producing the extended \Lya halos commonly observed around galaxies and tracing the structure and kinematics of their circumgalactic medium (CGM) \citep[e.g.,][]{steidel_structure_2010,steidel_diffuse_2011,wisotzki_extended_2016,fab19,maja22a, maja22b,gonzalez23} as well as spatial variations in the emergent spectrum \citep{Leclercq2020A&A...635A..82L,Li2022,Erb2023ApJ...953..118E}.

Such a wide range of observable phenomena illustrates the diagnostic power of \Lya: its emergent signal is highly sensitive to several physical parameters of the gas it traverses, including HI column density \citep{adams_escape_1972, neufeld_transfer_1990}, gas kinematics \citep{bonilha_monte_1979, ahn_ly_2002}, dust content \citep{neufeld_escape_1991, laursen_ly_2009,seon20}, and spatial structure or fragmentation \citep{hansen_lyman_2006,laursen13,gronke_mirrors_2016,chang_radiative_2023}. In principle, these imprints can be decoded from observables, allowing \Lya to provide a uniquely powerful window into the physical conditions of the ISM and CGM. Yet, the very resonant nature that makes \Lya so informative, also complicates its interpretation, as it is still not fully clear how the gas properties shape the \Lya line. 

To interpret these complex profiles, theoretical modeling is necessary. Over the past two decades, significant progress has been made in modeling \Lya radiative transfer (RT) to interpret observed line profiles. Many studies have focused on relatively idealized, isotropic geometries, such as expanding spherical shells or homogeneous slabs, which have proven successful in reproducing observed spectra \cite[e.g.,][]{verhamme_3d_2006,dijkstra_ly_2006, yang_green_2016, gronke_modeling_2017, song_ly_2020}. However, the physical meaning of fitted model parameters (e.g., HI column density or outflow velocity) often remains ambiguous, particularly given the simplified assumptions underlying these models \citep{orlitova_puzzling_2018, li_deciphering_2022}. In reality, the ISM and CGM are expected to be highly anisotropic, structured by turbulence, inflows, and feedback-driven outflows, creating porous, multiphase gas, and anisotropic distributions \citep{cox_large-scale_1974, mckee_theory_1977, faucher-giguere_key_2023}. This complexity raises a fundamental and still unresolved question: What does \Lya truly probe in a multiphase, anisotropic medium? Does the emergent spectrum reflect the average column density of HI, or is it biased toward the lowest-density sightlines that enable escape? 

Recent work has begun to explore \Lya RT through anisotropic configurations \citep[e.g.,][]{behrens_inclination_2014,smith_physics_2022, kimm_systematic_2022, Byrohl2023}, highlighting how deviations from isotropy can significantly affect \Lya observables and contribute to the diversity in line profiles as a result of galaxy orientation and evolutionary stage \citep{verhamme2012,blaizot_simulating_2023}. A particularly intriguing suggestion emerging from the assumption of anisotropy is that \Lya photons escape preferentially through low-density channels in the neutral gas, paths of least resistance carved by stellar feedback or galactic dynamics \citep{zackrisson_spectral_2013, jaskot_new_2019,kakiichi_lyman_2021}. To explore this idea quantitatively, \citeauthor{almadamonter_crossing_2024} (\citeyear{almadamonter_crossing_2024}; hereafter \citetalias{almadamonter_crossing_2024}) examined \Lya radiative transfer in a static slab of neutral hydrogen perforated with an empty hole. The naive expectation was that the line profile would emerge as a single peak at line center since photons are much more likely to escape through the empty hole than to traverse the optically thick slab. However, the results revealed a more complex behavior: increased scattering shifts the photon frequencies enough to escape \citep{adams_escape_1972}, leading to a higher transmission probability than predicted \citep{neufeld_transfer_1990}. Consequently, the emergent \Lya profile shows not only the central peak (from photons escaping through the hole) but also pronounced red and blue peaks from photons transmitted through the dense slab. These findings have important implications for how \Lya traces the structure and column density of neutral gas. The ability of \Lya photons to escape both through low-density holes and high-opacity regions suggests that \Lya may be sensitive to both extremes of the HI distribution, challenging the common assumption that \Lya escape is dominated solely by the paths of least resistance.

These insights into \Lya escape naturally lead to its potential as a tracer of Lyman continuum (LyC) photons, which are affected by the same neutral gas channels \citep{dijkstra_lyman_2014,hayes_lyman_2015,verhamme_using_2015,verhamme_lyman-_2017}. Observationally, measurements of \Lya such as the peak separation ($v_{\rm sep}$) and equivalent width (EW), show strong correlations with the ionizing escape fraction $f_{\rm esc}$ in local analogs of reionization-era galaxies. In particular, an anti-correlation is observed between $f_{\rm esc}$ and $v_{\rm sep}$, indicating that galaxies with smaller \Lya peak separations tend to have higher LyC leakage \citep{izotov_detection_2016,izotov_low-redshift_2018,izotov_lyman_2021,flury_low-redshift_2022-1,flury_low-redshift_2022}. However, given that these diagnostics have been studied theoretically in using isotropic gas distributions, it is unclear how they perform in more realistic geometries. Specifically, are the gas parameters probed by \Lya the average properties, weighted towards low optical depth pathways, or simply the line-of-sight properties? 
With respect to using \Lya as a proxy for LyC leakage, this  ties into the question whether \Lya probes the `global' LyC escape fraction, i.e., what LyC flux leaks into the IGM -- or the line-of-sight one. As the former governs the evolution of the intergalactic medium during the epoch of reionization but the latter is directly observable -- and the two can be vastly different \citep{Fernandez2010,Paardekooper2015,Ma2015} -- understanding which of these \Lya encodes is essential for interpreting observations and for constraining the sources of cosmic reionization.

In this paper, we investigate how the presence of an empty or low-column density channel affects the \Lya line profile, accounting for a range of gas properties and geometries (including single and multiple channels, varying column densities, outflows, and dust) as well as a more complex column density ($N_{\mathrm{HI}}$) distribution, expected in the ISM. In \S\ref{sec:numerical_setup}, we describe our Monte Carlo radiative transfer method and introduce our initial model setup, which forms the basis for the results presented in \S\ref{sec:results}. We discuss the broader implications of our findings in \S\ref{sec:discussion}. Finally, we summarize and present our conclusions in \S\ref{sec:conclusions}.

\section{Methods}
\label{sec:numerical_setup}
To explore the escape of \Lya from porous gas distributions, we perform Monte Carlo simulations through slabs containing low-column-density or empty channels that can act as a direct escape route. We describe the numerical method in \S~\ref{sec:monte_carlo}, then introduce the scattering geometries  in \S~\ref{sec:geometry}, followed by an analytical perspective on \Lya transfer through said geometries in \S~\ref{sec:analytic_view}

\subsection{Monte Carlo \texorpdfstring{\Lya}{Lya} radiative transfer}\label{sec:monte_carlo}

We use the Monte Carlo radiative transfer code \texttt{tlac} \citep{gronke_directional_2014} to simulate the propagation of \Lya photons through a neutral dusty hydrogen medium. The photon frequency is expressed using the dimensionless form $x$, which is given by 
\begin{equation}
    x=\frac{\nu-\nu_0}{\Delta \nu_D},
\end{equation}
where $\nu_0=2.46\times10^{15}$ Hz is the line-center frequency and $\Delta \nu_D\approx 1.1\times10^{11} (T/10^{4} \ \mathrm{K})^{1/2}$ Hz is the Doppler thermal line broadening.  Negative (positive) values of $x$ correspond to redward (blueward) frequencies relative to line center.

As a first step, each photon is initialized with frequency at the line center ($x=0$) and a propagation direction sampled uniformly across all directions. Then an optical depth $\tau$ is sampled from the exponential distribution $\exp(-\tau)$. This optical depth is converted into the physical distance $\ell$ the photon travels before interacting with a hydrogen atom or a dust grain via
\begin{equation}
    \tau=\int^{\ell}_0 \left[\sigma_{\mathrm{HI}}(x,s) n_{\mathrm{HI}}(s)+\sigma_{\rm d}(s) n_{\rm d}(s)\right]\mathrm{d}s,
\end{equation}
with the interaction type determined probabilistically by the local abundances of both dust and hydrogen. Here, $n_X$ and $\sigma_X$ denote the corresponding number density and cross section. 
The \HI cross section $\sigma_{\rm HI}(x,s)$ follows the Voigt–Hjerting function and depends strongly on frequency, whereas the dust extinction cross section $\sigma_{\rm d}(s)$ is nearly frequency-independent.

If the photon scatters with hydrogen, its new direction is drawn from the resonant phase function (nearly isotropic in the line core, and increasingly dipole-like in the line wings). If dust interaction occurs, the photon is either absorbed or scattered depending on the dust albedo $A_{\rm d}=0.32$ \citep[from the Milky way dust model proposed by][]{li_infrared_2001}. In case of dust scattering, the new direction is chosen to follow the \cite{henyey_diffuse_1941} phase function with an asymmetry parameter of $g_{\mathrm{a}}=0.73$.

The code follows individual photon packages through position and frequency space until they escape or are absorbed. The emergent spectrum is then constructed from the frequencies of photons that escape the simulation volume.

\subsection{Scattering geometries and numerical method}\label{sec:geometry}

\begin{figure}
    \centering   
    \includegraphics[width=0.97\linewidth]{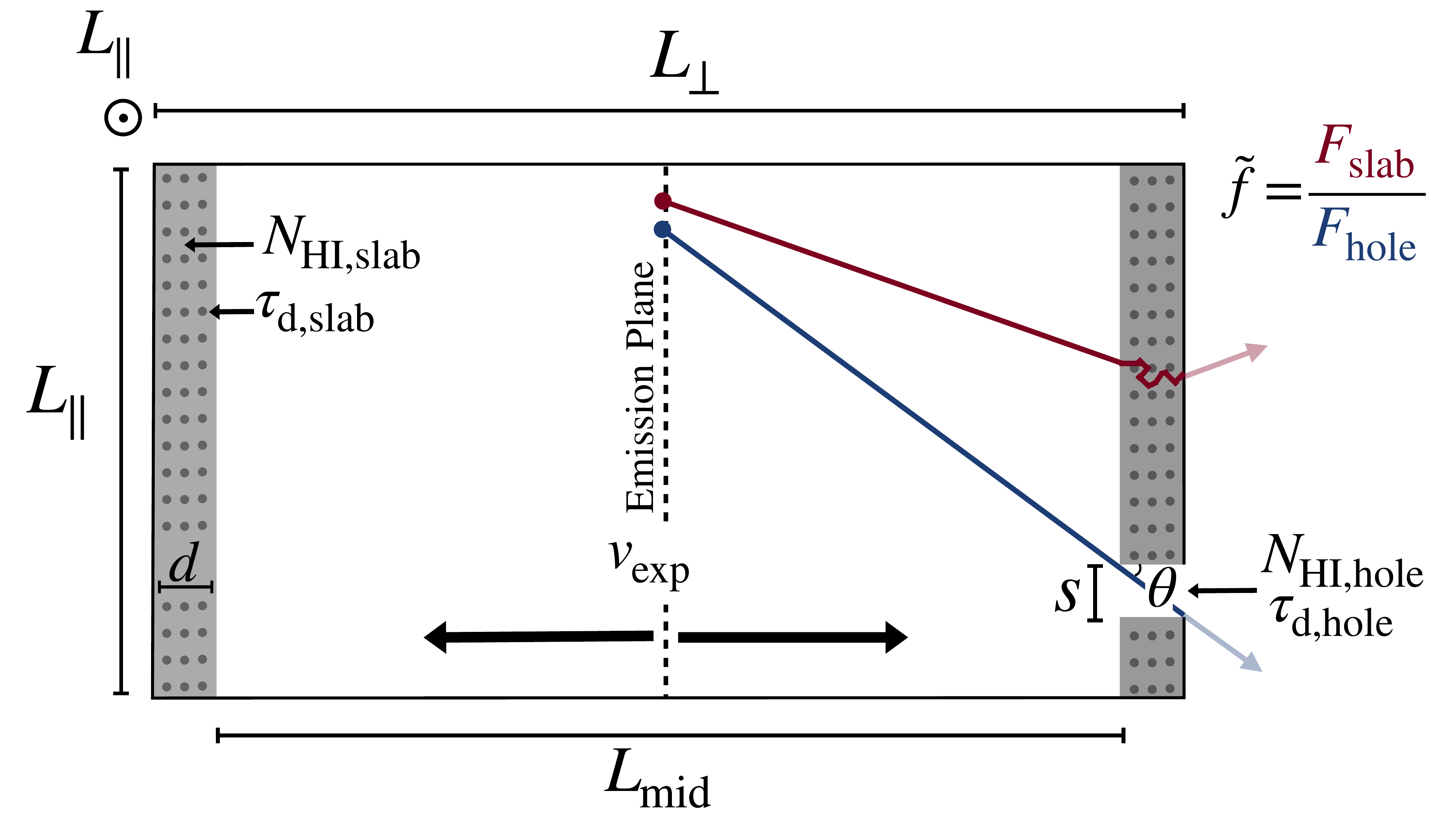}
    \caption{Simulation setup to study the effect of non-uniform gas distributions on the \Lya profile. The box has periodic boundaries along the $\parallel$-axes and contains a slab with a thin layer of neutral gas and dust. Anisotropy is introduced by piercing the gas layer with a square hole of side length $s$ along the non-periodic axis. The arrows emerging from the emission plane trace the trajectories of photons escaping either through the slab or through the hole at the maximum angle $\theta$, set by the width $d$.}
\label{fig:diagram}
\end{figure}

We performed Monte Carlo radiative transfer simulations in a semi-infinite slab system, with the emission plane located at the center. The box dimensions are $L_{\parallel}^2 \times L_{\perp}$. Here, the $\parallel$- and $\perp$-directions are defined relative to the emission plane. The emission plane separates the box into two thin neutral hydrogen layers, each of width $d$ (and a corresponding column density $N_{\rm HI,slab}$), and a central empty region of width $L_{\mathrm{mid}}$ (i.e., $L_{\perp} = L_{\mathrm{mid}} + 2d$). Periodic boundary conditions were applied along both the $\parallel$-axes (see Figure~\ref{fig:diagram}).

To explore how non-uniform gas distributions affect the \Lya line profile and its escape, we introduced a lower density region within the slab: a square hole of side $s$ that pierces completely through one of the slabs. Typically, we choose this hole to be empty, i.e., with \HI column density $N_{\rm HI,hole}=0$, but we also vary this value (see below). The hole defines an effective fractional opening area of \HI slab $\tilde s$ is given by
\begin{equation}
\tilde s = s^2 / L_{\parallel}^2,
\end{equation}
where $s^2$ and $L_{\parallel}^2$ represent the area of the hole and the whole slab, respectively.

In addition to this baseline setup, we explored several modifications to assess the sensitivity of the \Lya spectrum to physical and geometric variations. Each of these variations was tested in isolation to examine its individual effect on \Lya radiative transfer. These included: 
\begin{enumerate}
    \item Increasing the width of the slab $d$ (see Fig.~\ref{fig:diagram}), forming a tunnel-like channel to study its influence on \Lya escape. \label{item:doortunnel}
    
    \item Subdividing a given hole of area $\tilde s$ into $n_{\mathrm{h}}$ smaller holes of side $s_i=s/\sqrt{n_{\mathrm{h}}}$ distributed randomly to test the impact of hole area distribution. This mimics changing not only the presence of low-density channels but also varying the overall porosity of the ISM.\label{item:porosity}
    
    \item Allowing the hole to be filled with gas of a lower \HI column density $N_{\mathrm{HI,{hole}}}$ than the surrounding slab with $N_{\mathrm{HI,{slab}}}$, which can be due to, e.g., a fiducial residual neutral fraction, within an otherwise ionized channel.
    
    \item Introducing a constant outflowing velocity $v_{\mathrm{exp}}$ in the gas, orthogonal to the emission plane. Since \Lya photons are highly susceptible to bulk flows produced, for instance, by ISM feedback mechanisms, this case is important to understand the impact of the outflow on \Lya escape and the asymmetric observed profile. 
    
    \item Adding dust with an optical depth $\tau_{\mathrm{d, slab}}$ in the slab, and $\tau_{\mathrm{d, hole}}$ in the hole. As the dust in the ISM causes the extinction of UV radiation (i.e, \Lya), it is essential to understand how \Lya is formed as it escapes through dusty ISM. \label{item:dust}
    
    \item Generating slabs with log-normally distributed column densities, fixing either the mean, median, or mode at a specified value. 
    The distributions of column densities are characterized by the deviation $\sigma$; the mean, median, and mode distributions are $\propto \exp(\mu+\sigma^2/2)$, $\exp(\mu)$, and $\exp(\mu-\sigma^2)$, respectively.
    \label{item:lognorm} 
\end{enumerate}
All slabs contain static neutral hydrogen with a column density $N_{\mathrm{HI}}$ between $10^{17}$ and $10^{20}$ $\mathrm{cm}^{-2}$ at a fixed temperature $T=10^4$ K (see Appendix~\ref{sec:largeatau0} for considerations applying to lower temperatures). The column density is uniform across the slabs in all simulations except for those in \ref{item:lognorm}, where it is assigned according to the log-normal distribution.

For all simulations except for those in \ref{item:doortunnel} and \ref{item:porosity}, the slab thickness is extremely thin ($d\ll L_{\perp}$) such that the first point of interaction is not dependent on the initial position of the photon (as is also the case, for a spherical setup). To resolve the channels/holes, the domain of the parallel sides of the grid is divided into 100 cells each. This resolution allows us to study holes with a minimum effective opening area of $\tilde{s}=0.01$. On the other hand, to resolve the extremely thin layer of neutral gas, the perpendicular domain requires a much higher resolution. We adopt 50 000 cells, which is sufficient to resolve the HI thickness.
Finally, we initialize $10^5$ photon packets at the emission plane to track their propagation through the box. In simulations that include dust \ref{item:dust}, we increase the number of photon packets to $10^6$ to compensate for photon losses due to dust absorption.
Specifically, we track numerically if they escape through the hole or through the slab, yielding the flux ratio of escape pathways. Numerically, we classify a photon as escaping through the hole if the column density it passed through is $<N_{\rm HI,slab}$; otherwise, it is considered to escape directly through the slab.

While the geometry described above is simplified, it represents the essence of a range of astrophysical environments. For example, it can represent situations in which ionizing radiation escapes through low–column-density channels filled by residual neutral hydrogen with $N_{\rm HI}\lesssim 10^{17} \cm^{-2}$, embedded within gas more than 3 orders of magnitude larger column densities, as suggested by theoretical, numerical and observational studies \citep[e.g.,][]{kakiichi_lyman_2021, Ma2015, gazagnes_origin_2020}. Similar configurations may also arise on larger scales, such as in ISM “blow-out” chimneys carved by stellar-driven feedback and galactic winds \citep[e.g.,][]{HeckmanThomson2017,Fujita2003,Kimm2019}. 

In addition, more complex morphologies found in radiation-hydrodynamic simulations \citep[e.g.,][]{mitchell_tracing_2021,blaizot_simulating_2023, Bhagwat_Lya_2025} are regularly difficult to interpret directly in terms of \Lya observables, making idealized models, such as those explored in this work, a useful framework for isolating the key physical processes at play. Idealized models, such as those explored in this work, therefore provide a useful bridge between simulations and the physical processes that determine the emergent \Lya radiation.

\subsection{Analytical considerations}\label{sec:analytic_view}
In \citetalias{almadamonter_crossing_2024}, we introduced the hole-to-slab flux ratio $\tilde{f}$ to quantify the relative contribution of photons escaping through a low-density channel (the `hole') versus those transmitted through a surrounding static, optically thick medium 
\begin{equation}
    \tilde f=\frac{F_{\mathrm{hole}}}{F_{\mathrm{slab}}}.
\end{equation}

Here, we present an alternative derivation of the fraction of \Lya photons expected to escape through the low-density `hole' (with fractional size $\tilde s$) versus the ones piercing through the slab.

While the number of \textit{reflections} required in order to escape through the hole is simply given by $N_{\rm hole}^{\rm refl}\approx 2/\tilde s$, the number of \textit{scatterings} required to escape through the slab is $N_{\rm slab}^{\rm sctr}=T_{\rm slab}^{-1}$ where $T_{\rm slab}$ is the transmission probability of the slab. Thus, using the mean number of scatterings per reflection $\bar{N}_{\rm refl}^{\rm scat}$, one can express the flux fraction escaping through the hole as
\begin{equation}
\tilde f\approx \frac{N^{\rm scat}_{\rm slab} / \bar N_{\rm refl}^{\rm scat}}{N^{\rm refl}_{\rm hole}}\approx \frac{\tilde s / 2}{T_{\rm slab}\bar{N}_{\rm refl}^{\rm scat}}.
\label{eq:ftilde1}
\end{equation}

In order to evaluate  $\bar{N}_{\rm refl}^{\rm scat}$, we can consider a \Lya photon penetrating $\sim \lambda_{\rm mfp}$ into the  slab. It has a $\sim 50\%$ probability to exit the slab again, i.e., be reflected, and $\sim 50\%$ to penetrate $\sim \lambda_{\rm mpf}$ deeper into the slab.
 This problem is analogous to the `Gambler's ruin' problem, where a gambler plays $n$ games and starts with an initial wealth of $w_0$. The gambler can win or loose one dollar at each game\footnote{Note that we will only consider one absorbing barrier (at $w=0$) and not two. This is because the photon escapes after a fixed number of games/scatterings and not at a specific point in space. See Appendix~\ref{sec:largeatau0} for a generalization.}. It turns out that while the probability of eventually losing all the money is unity, the average number of games required to do so is infinity.
 
 In our case, a \Lya photon has penetrated a length of $w_0 = l_0 / \lambda_{\rm mfp} \approx 1$ into the slab, and has an equal probability ($p=0.5$) to penetrate a distance of $\sim \lambda_{\rm mfp}$ deeper into -- or out of -- the slab. Thus, while the naive expectation of a large overall reflection probability is supported, the number of scatterings per reflection is large. This is crucial as every scattering implies a small probability of frequency shift to $x\ge x_{\rm esc}$ at which the optical depth is so low that escape is possible \citep{adams_escape_1972,neufeld_transfer_1990}\footnote{The analogy for the Gambler's ruin would be a fixed sum of money at which the Gambler leaves the casino; thus lowering the probability of loosing all their money.}.

The distribution of distances after $n$ steps is therefore a shifted and rescaled binomial distribution $b(w|n,p)$
\begin{equation}
    f(w|n) = b(\frac{1}{2}(w + n - w_0)|n,p=\frac{1}{2}).
\end{equation}
After $n$ steps, the distribution of steps of \Lya that are reflected is given by 
\begin{equation}
    f_{\rm reflected}(w|n) = \begin{cases}
    f(w|n) \text{ for } w\le 0\\
    f(-w|n) \text{ for } w>0
    \end{cases}
\end{equation}
where the former is clear (those photons have exited the slab already) and the latter comes from the `reflection principle', that is, for every photon that has a distance $w<0$ after $n$ steps, there is exactly one (because of mirrored trajectories) that has $w>0$.

Now the fraction of photons that are reflected after $n$ steps is the cumulative
\begin{align}
    F_{\rm reflected}(n) =& \sum\limits_{w=-\infty}^{\infty} f_{\rm reflected}(w|n)\\
    =& F(0|n) + F(-1|n)\\
    =& B(\frac{n-w_0}{2}|n) + B(\frac{n-w_0}{2} - 1|n)
\end{align}
where $F$ and $B$ are the cumulative distribution functions  (CDFs) to $f(w|n)$ and the binomial distribution, respectively\footnote{One technicality is that since in this example the photons require $w_0 + 2 m$ steps to be reflected ($w_0$ for a straight walk to ruin and detours must come in pairs), $F_{\rm reflected}(n)$ is only well defined 
for $(n+w_0)/2 \in \mathbb{N}$.}.

This implies the number of photons that are reflected at step $n$ is given by
\begin{align}
    f_{\rm reflected}(n) =& F_{\rm reflected}(n) - F_{\rm reflected}(n-2).
\end{align}

Writing out $f_{\rm reflected}(n)$ using Binomial coefficients and using the Stirling's approximation ($n \approx \sqrt{2 \pi n}(n/e)^n$), one finds that for large $n$, $f_{\rm reflected}\propto n^{-3/2}$ (we checked this explicitly in \citetalias{almadamonter_crossing_2024}). We can use this to compute the mean number of scatterings per reflection by evaluating
\begin{equation}
 \bar N_{\rm refl}^{\rm scat} = \sum\limits_{n = 1}^{C_1 \tau_0} f_{\rm reflected}(n) n \approx \sqrt{C_2 \tau_0}\;.    
\end{equation}
Here, we used $C_1 \tau_0$ (with $C_1 \approx 2.66$) as a cutoff, as this is the average number of scatterings required in order to shift to a frequency sufficient for escape \citep{adams_escape_1972}. Taking into account the proper constant prefactor for $f_{\rm reflected}$, one finds $C_2=(2 C_1/\pi)^{1/2}$ consistent with the continuous version \citepalias{almadamonter_crossing_2024}. 

In addition, one can find from $f_{\rm reflected}$ the transmission probability through the slab
\begin{equation}
    T_{\rm slab}=\alpha \sum\limits_{n=C_1\tau_0}^{\infty}f_{\rm reflected}(n)\approx \alpha \zeta\left(\frac{3}{2},C_1\tau_0\right) \sim \tau_0^{-1/2}
    \label{eq:transmission}
\end{equation}
where $\alpha \approx 1/2$ is a numerical prefactor due to the fact that once photons reach $x\gtrsim x_{\rm esc}$ there is an equal chance of escaping through the slab on either side (see Appendix~\ref{sec:largeatau0}, the case of a general $\alpha$). We furthermore used $f_{\rm reflected}\propto n^{3/2}$ for the first approximation, and $\zeta$ here is the Hurwitz zeta function, which is to leading order for large $\tau_0$ approximately $\tau_0^{-1/2}$. 

Plugging in $T_{\rm slab}$ and $\bar N_{\rm refl}^{\rm scat}$ in Eq.~\ref{eq:ftilde1} yields as a hole-to-slab ratio
\begin{equation}
    \tilde f\approx \epsilon \frac{\pi}{2}\tilde s
    \label{eq:ftilde}
\end{equation}
where $\epsilon\approx 5$ is a fudge factor which contains the numerical prefactors from above. 

Note \textit{(i)} $\tilde f\sim \tilde s$, i.e., the fraction of photons transmitted through the (optically much thicker) slab is relatively small, and \textit{(ii)} that the dependencies on $\tau_0$ canceled each other which means the fraction of photons escaping through the channels is independent of the optical depth of the surrounding slab. 
While both of these facts seem counterintuitive at first, one needs to consider that an optically thicker medium means that it is harder to transmit through it, but it is also `stickier', that is, the number of scatterings -- and thus for \Lya the probability to shift to an escape frequency -- increases. Thus, the mental image we should have in mind when thinking about \Lya scattering and escape is not ping-pong balls bouncing off some optically thick walls but rather a hiker stepping into patches of dense forest: once inside, they wander around between many trees before eventually stumbling back out\footnote{An analogy closer to radiation is light entering frosted glass which scatters many times internally before exiting}.

\section{Results}
\label{sec:results}

\subsection{Door vs Hallway}\label{sec:door_hallway}
A hole carved on an extremely thin layer of gas ($d\ll L_\perp$) provides a helpful starting point. However, low-column-density channels may also take the form of elongated, tunnel-like structures shaped by stellar feedback and inhomogeneous ISM distributions. In this section, we study the effect of \Lya escape along extended, low-density paths. This allows us to assess how the spatial extent and shape of low-opacity regions influence \Lya radiative transfer and the resulting spectral features.

The hole-slab ratio of the escaping flux, $\tilde{f}$, depends on the probability that photons find the hole after each time they are reflected. As discussed in \S~\ref{sec:analytic_view}, in the case of an extremely thin slab (i.e.,  for a small $d$), this probability is simply $\tilde s/2$, where the factor $1/2$ arises from the fact that half of the photons are directed towards the slab containing the hole, while the remaining half are directed towards the slab without the hole. Under these conditions, it is reasonable to assume negligible interaction between photons and the thin walls surrounding the hole. However, as the gas layer thickness increases, the hole effectively transforms into a tunnel-like structure, acting as a corridor where photons can be absorbed by its walls, thereby reducing the likelihood of escape. Consequently, the probability that a photon encounters the hole must be modified by a geometrical factor $g \leq 1$, which depends on the thickness of the neutral hydrogen layer $d$ (see Fig. \ref{fig:diagram}).

We can estimate the geometrical factor $g$ by considering all possible emission angles $\beta$ (defined from the orthogonal of the emission plane, cf. Fig.~\ref{fig:diagram}) at which the `thin slab approximation' is true. This is the case if a photon cannot hit the walls of a hole, i.e., if $|\beta| < \theta = \arctan(\sqrt{2}s/ d)$ (cf. Fig.~\ref{fig:diagram}, using $\sqrt{2}$ as a geometrical prefactor). This range defines a solid angle of $\Omega_g=2 \pi [1-\cos(\theta)]$ or a prefactor
\begin{equation}
    g = \frac{\Omega_g}{2 \pi} = 1 - \frac{d}{\sqrt{d^2 + 2 s^2}}
\end{equation}
which alters the expected flux escaping through the hole (from Eq.~\eqref{eq:ftilde}) to
\begin{equation}
\tilde{f} \approx\frac{g\pi}{2} \tilde s.
\label{eq:ftildeg}
\end{equation}

Note that in the limit of an extremely thin gas layer $d\rightarrow 0$ then $\mathrm{g} \rightarrow 1$, recovering Eq.~\eqref{eq:ftilde}.

\begin{figure}
    \centering
\includegraphics[width=\linewidth]{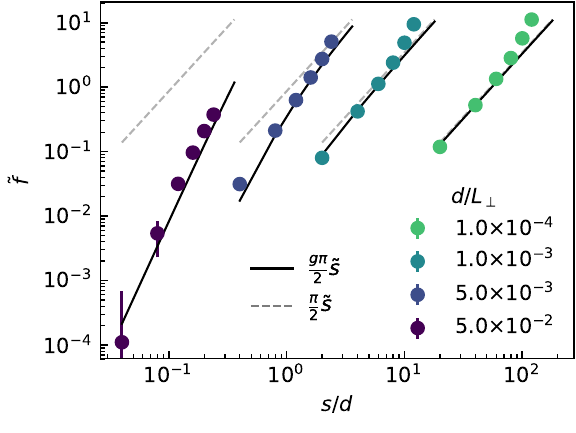}
\vspace{-0.5cm}\caption{Hole-slab flux ratio $\tilde f$ as a function of the hole size $\tilde s$ to tunnel depth $\rm d$.Points on the left correspond to the “hallway” regime, gradually transitioning to the “door” regime toward the right. The column density is fixed at $N_{\rm HI} = 10^{19}\ \mathrm{cm^{-2}}$. Simulated results closely follow the analytical prediction for $\tilde{f}$ when the geometry factor $\mathrm{g}$ is included (solid lines). For comparison, we also show $\tilde{f}$ without applying the geometry correction (dashed lines).}
    \label{fig:ftilde_g}
\end{figure}

We tested Eq.~\eqref{eq:ftildeg} using numerical simulations, as illustrated in Figure~\ref{fig:ftilde_g}. In these simulations, we fixed the value of $d$ and varied the hole size $s$, assuming a column density of $N_{\mathrm{HI}} = 10^{19}\,\mathrm{cm}^{-2}$. To highlight the impact of the correction with $g$, we plot both the corrected (Eq.~\eqref{eq:ftilde}) and uncorrected ($g$=1 for all $d$ cases with a solid and dashed line, respectively. The results show that Eq.~\eqref{eq:ftilde} provides a good fit to the numerical simulations under these conditions. Note that, when a photon encounters the ``walls'' of the hallway, which is more likely for larger values of $d$, the hole-to-slab ratio $\tilde{f}$ decreases. This occurs because, as discussed in \citetalias{almadamonter_crossing_2024}, photons are more likely to be absorbed by the walls and undergo multiple scatterings before reflection. During this process, the photon's direction likely changes enough to prevent it from re-entering the hallway and from exiting through the tunnel.

\subsection{Photon Escape Direction}
\label{sec:appendix_escape_direction}
\begin{figure}
    \centering
    \includegraphics[width=1\linewidth]{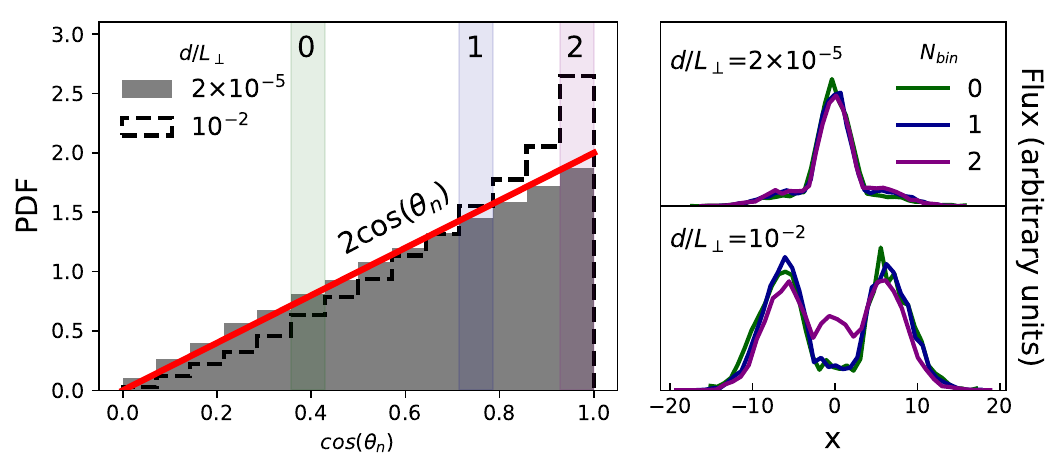}
    \caption{\textit{Left}: Distributions of the photon escape angle $\theta_n$ for the thin (`door') and the thick (`hallway') hydrogen layer configurations. The solid line represents the expected behavior of escape directions from locally isotropic emission. \textit{Right:} \Lya line profiles coming from different escape angle bins (also marked in the left panel).}
    \label{fig:not_beamed}
\end{figure}

The left panel of Figure \ref{fig:not_beamed} shows the probability density function of the escape angle $\theta_n$, measured with respect to the slab's normal, for all photons escaping from both the plane and the hole. The parameter $d/L_\perp$ with $d$ and $L_\perp$ as shown in Fig. \ref{fig:diagram} quantifies the thickness of the neutral hydrogen layer. In red, we also show the expected behavior of escape directions from a Lambertian plane, corresponding to locally isotropic emission, i.e., $dP/d\Omega\propto{\cos(\theta_n)}$, resulting in $p(\mu)=2\mu$ with $\mu=\cos(\theta_n)$. Note that for the lowest value of $d/L_\perp$, i.e., the ``door'' geometry (or the thin hydrogen layer configuration, as described in \S~\ref{sec:door_hallway}), the escape angles of photons are very close to local isotropy and show no evidence of strong beaming. When the hole is embedded within a thick neutral hydrogen layer (i.e., the larger $d/L_\perp$), photons escape through what we refer to as a ``hallway'' geometry. In this case, escape becomes mildly angle-dependent, since only photons within a limited range of directions can exit through the channel without finding the inner walls of the hallway. This results in slightly beamed escape. 

Furthermore, the colored stripes in the left panel of Fig.~\ref{fig:not_beamed} correspond to bins in escape angle, each associated with a \Lya line profile with the same color and number on the left panel. Angular bins that follow a locally isotropic escape distribution, show very similar line profiles, indicating the absence of strong angular dependence in the spectrum. The largest deviation from this behavior occurs in bin 3 for the case of a thicker neutral hydrogen layer ($d/L_\perp=10^{-2}$), where the line profile differs from the rest of the spectra. This deviation reflects the mild angular dependence that arises in the ``hallway'' scenario.

Despite the mild angular dependence in the runs with a larger $d/L_\perp$, we find that, in general, the escape is much less beamed than naively anticipated\footnote{This had been previously concluded in \citet{behrens_beamed_2014}. However, they focused on the low-optical depth (mostly large outflow) regime.}. The initial expectation was that photons would preferentially escape through the hole, which in the case of the thick hydrogen layer, implies that the escape would be strongly beamed because of the limited directions allowed by the hole. Instead, we find a wide range of escape angles thanks to the non-negligible fraction of photons escaping from the surrounding slab with no preferred direction. 

\subsection{Porosity}
\begin{figure*}
    \centering
    \includegraphics[width=\linewidth]{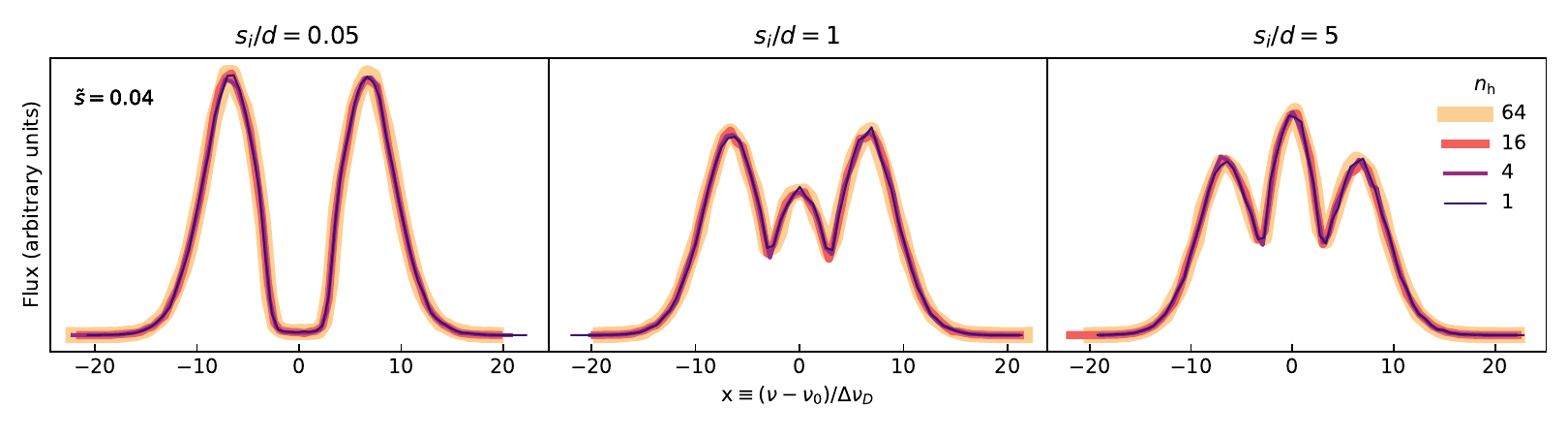}\vspace{-0.25cm}\caption{Emergent \Lya spectra from slab systems with varying hole geometric configurations. In each panel, the ratio between individual hole size $\bar{s_i}$ and tunnel depth $\rm d$ is held constant, while the total hole area is fixed at $\tilde s = 0.04$ and distributed equally across different numbers of holes. Solid lines of different thickness indicate the number of holes. The spectra show that as long as the ratio $\bar{s_i}/d$ remains constant, the number of holes does not significantly affect the \Lya line profile.}\label{fig:porosity_spec}
\end{figure*}

In realistic astrophysical environments, low-column density paths through the interstellar medium are unlikely to resemble a single, coherent opening. Instead, what may appear as a single hole in simplified models likely consists of multiple, spatially separated low-opacity regions, effectively a collection of smaller sub-holes. These substructures can result from turbulent flows, fragmented outflows, or localized feedback-driven clearing, and their spatial distribution and covering fraction can significantly influence \Lya radiative transfer. In this section, we extend our analysis by replacing the idealized single-hole configuration with multiple sub-holes of equivalent total area. 

We explored the effect of subdividing the original hole of area $\tilde s$ into multiple smaller sub-holes, randomly distributed across the same projected area of the slab. Specifically, we split the hole into 4, 16, and 64 equal-area components. To incorporate this into our geometric model, we modify the angular parameter $\theta$ in \S~\ref{sec:door_hallway} to account for the angle required for photons to escape through the smaller individual holes:
\begin{equation}\label{eq:theta_si}
\theta= \arctan\left(\frac{\sqrt{{2}}s_i}{d}\right),
\end{equation}

where $s_i=s/\sqrt{n_{\mathrm{h}}}$ is the side of the individual hole and $n_{\rm h}$ is the number of holes. Whether the photon escapes via a ``door'' or ``hallway'' regime depends on the ratio $s_i/d$, as discussed in \S\ref{sec:door_hallway}.

Figure~\ref{fig:porosity_spec} shows the resulting spectra for different hole subdivisions, keeping the total hole area fixed at $\tilde s = 0.04$ and maintaining a constant ratio between the hole's side $s_i$ and hole depth $d$. Here, the number of holes (marked with different colors and sizes) does not strongly affect the \Lya line profile. As long as the geometry stays within the same regime (either `door' or `hallway'), splitting the area into more holes doesn't make a difference in the shape of the \Lya line profile. 

\begin{figure}
    \centering
    \includegraphics[width=\linewidth]{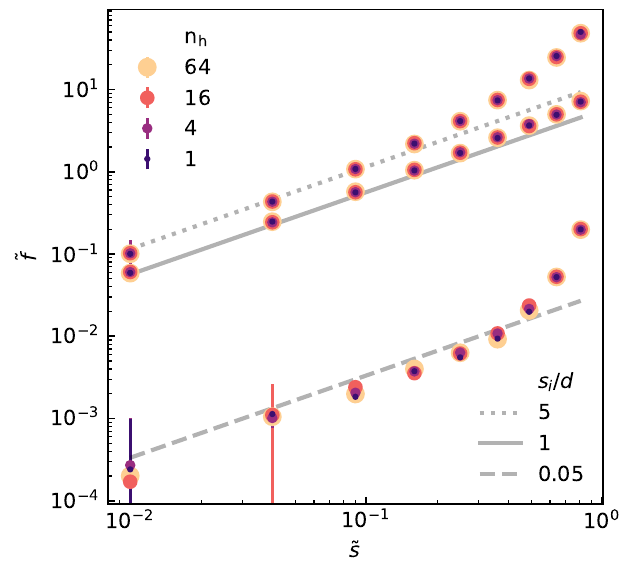}
    \vspace{-0.5cm}
    \caption{Hole-slab flux ratio $\tilde f$ as a function of total hole size $\tilde s$ keeping a constant ratio of the sub-hole individual size $\bar{s_i}$ and the tunnel depth $\rm d$. The number of holes, marked in the figure with different marker sizes, does not change the behavior of $\tilde f$ significantly. Predicted fluxes are marked with different line styles.}
    \label{fig:ftilde_porosity}
\end{figure}

Figure~\ref{fig:ftilde_porosity} compares the hole-slab ratio $\tilde{f}$ as a function of the total hole area $\tilde s$, for different values of $s_i/d$, together with a theoretical prediction from Eqs.~\eqref{eq:ftildeg} and ~\eqref{eq:theta_si}. Consistent with Figure~\ref{fig:porosity_spec}, the number of holes does not affect $\tilde{f}$ as long as the ratio between individual hole size and tunnel depth remains constant (that is, as long as the hole remains in either the `tunnel' or `door' regime). 

\subsection{Filled Channel}
\label{sec:filled_hole}
So far, we have considered empty escape channels or holes through which photons could travel without interacting with any material. However, in realistic astrophysical environments, a completely empty channel is unlikely. Processes such as gravitational inflows, residual neutral gas left over from incomplete feedback, or partial collapse of surrounding material can leave behind low-density hydrogen within these channels. As a result, even holes may contain enough neutral hydrogen to affect the transport of  \Lya (and potentially LyC)  photons. In this section, we explore how introducing gas into the channels affects the transport of radiation. We filled the holes with column densities ranging from much lower to comparable with that of the slab to understand how channels filled with low-density neutral gas can shape the emergent spectra.

\begin{figure*}
\centering
\includegraphics[width=0.6\linewidth]{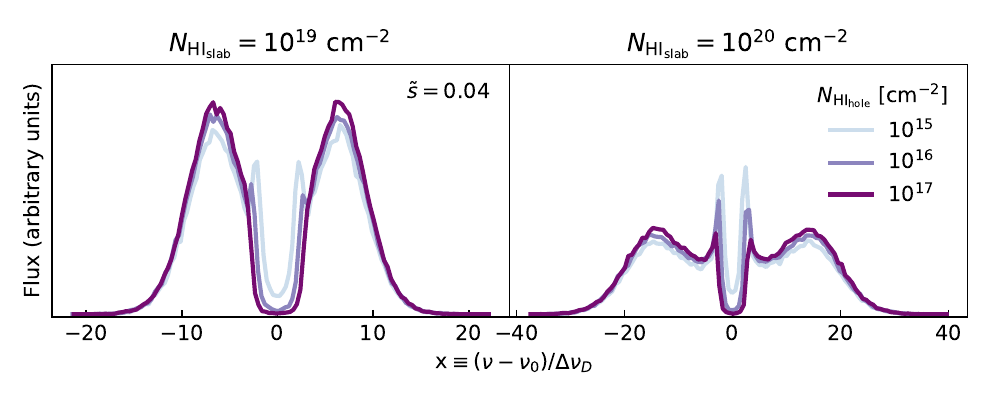}
\includegraphics[width=0.3\linewidth]{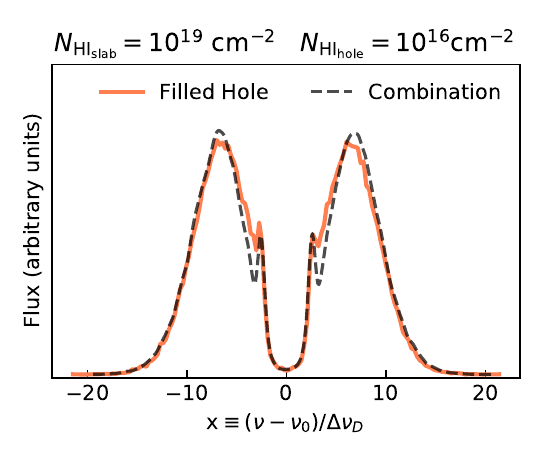}
\vspace{-0.5cm}
\caption{Emergent \Lya spectra from slab systems with optically thick hole ($N_{\rm HI,hole} > 10^{14}\, \rm cm^{-2}$). \textit{Left \& Center:} Effect of varying the column density inside a filled hole on the emergent \Lya spectra. \Lya photons simultaneously trace the high column density of the slab and the lower column density within the hole. When the contrast between the two is significant (more than two orders of magnitude), a distinct central peak emerges. For more minor differences, the inner emission blends with the broader slab features, often resulting in asymmetric red and blue peaks. \textit{Right}: Example of a \Lya spectrum from a slab with a filled hole. The hole is filled with gas at a column density of $N_{\rm HI_{Hole}} = 10^{16}\ \mathrm{cm^{-2}}$, while the surrounding slab has $N_{\rm HI_{Slab}} = 10^{19}\ \mathrm{cm^{-2}}$. The solid line shows the emergent spectrum from the slab with the filled hole. The dashed line represents the $\tilde f$ scaled combination of two separate slabs with the same respective column densities.}
    \label{fig:filled_comb_spec}
\end{figure*}

The left and central panels of Figure~\ref{fig:filled_comb_spec} show examples of emergent spectra from slab systems where the central hole has been filled with neutral hydrogen at various column densities $N_{\mathrm{HI,{hole}}}$, spanning up to five orders of magnitude below that of the surrounding slab, $N_{\mathrm{HI,{slab}}}$. These spectra exhibit the characteristic red and blue peaks resulting from photon scattering in the high-density static gas in the slab and, at the same time, a central flux component that traces the column density of the gas filling the hole. The peak separation must be sufficiently broad to distinguish the central feature clearly. When $N_{\mathrm{HI,{slab}}}$ and $N_{\mathrm{HI,{hole}}}$ differ by only one or two orders of magnitude, the inner red and blue components blend with the broader slab peaks, making them difficult to resolve and often resulting in asymmetric red and blue peaks (see, e.g., the spectra on the central panel of Fig.~\ref{fig:filled_comb_spec}). These results are consistent with the statement we discussed in \citetalias{almadamonter_crossing_2024}; namely, \Lya photons can trace both high and low column densities simultaneously.  

The first two panels of Figure~\ref{fig:filled_comb_spec} show an example of emergent spectra from a slab with a central hole of size $\tilde{s} = 0.04$, filled with neutral hydrogen at a column density of $N_{\mathrm{HI_{hole}}} = 10^{15}, 10^{16}$ and $10^{17}\,\mathrm{cm}^{-2}$, and embedded within a surrounding slab with $N_{\mathrm{HI_{slab}}} = 10^{19}$ and $10^{20}\cm^{-2}$. By combining and appropriately scaling the emergent spectra from two independent slab systems without holes (for two \HI column densities of $10^{19}$ and $10^{16}\cm^{-2}$, respectively in the case of Fig.~\ref{fig:filled_comb_spec}) using the corresponding factor $\tilde f$, it is possible to reproduce the positions of the main spectral peaks closely: four distinct peaks when the column densities differ by more than an order of magnitude, or two blended peaks when the difference is smaller. This simplified construction provides a useful first-order approximation to the full system. The resulting combined spectrum (smoothed with a Gaussian filter with width $\sigma_g = 2\kms$) is shown as a dashed line in the right panel of Figure~\ref{fig:filled_comb_spec}. Note that some radiative effects near the transition between the two components are not fully captured. These come from photons escaping through the hole after suffering a few scatterings in the back slab. Nevertheless, the combined spectrum offers a valuable approximation of the overall spectral morphology and highlights key radiative features shaped by this inhomogeneous configuration.

\begin{figure}
    \centering
    \includegraphics[width=0.95\linewidth]{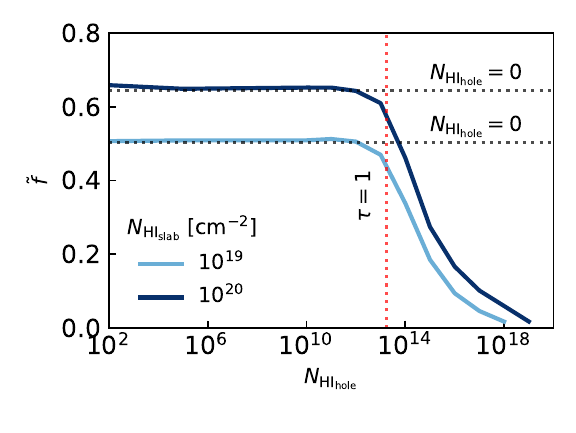}
    \caption{Hole-to-slab flux ratio $\tilde{f}$ as a function of the column density in the filled hole, $N_{\rm HI_{hole}}$. At low column densities, the filled hole behaves similarly to an empty one (dashed horizontal lines). As $N_{\rm HI_{hole}}$ increases and the hole becomes optically thick to \Lya photons (transition indicated by the vertical dotted line), scattering within the hole begins to alter photon paths, reducing their likelihood of escaping through the channel.}
    \label{fig:ftildevsfilled}
\end{figure}

Figure~\ref{fig:ftildevsfilled} shows the behavior of $\tilde{f}$ as a function of $N_{\rm HI,hole}$. When $N_{\mathrm{HI_{hole}}} \leq 2 \times 10^{13} \ \mathrm{cm^{-2}}$, the hole remains optically thin to \Lya photons, allowing them to escape with minimal interaction. In this regime, $\tilde{f}$ closely follows the trend of the empty-hole case (indicated by the dashed line in the figure). Once the column density exceeds this threshold, the hole becomes optically thick ($\tau > 1$), and photons undergo more scatterings within the hole, changing direction and hence reducing the likelihood of escape. Consequently, for $N_{\mathrm{HI_{hole}}} > 2 \times 10^{13} \ \mathrm{cm^{-2}}$, $\tilde{f}$ decreases with increasing $N_{\rm HI,hole}$.

\subsection{Outflows}
\label{sec:outflows}
\begin{figure*}
    \centering
    \includegraphics[width=0.9\linewidth]{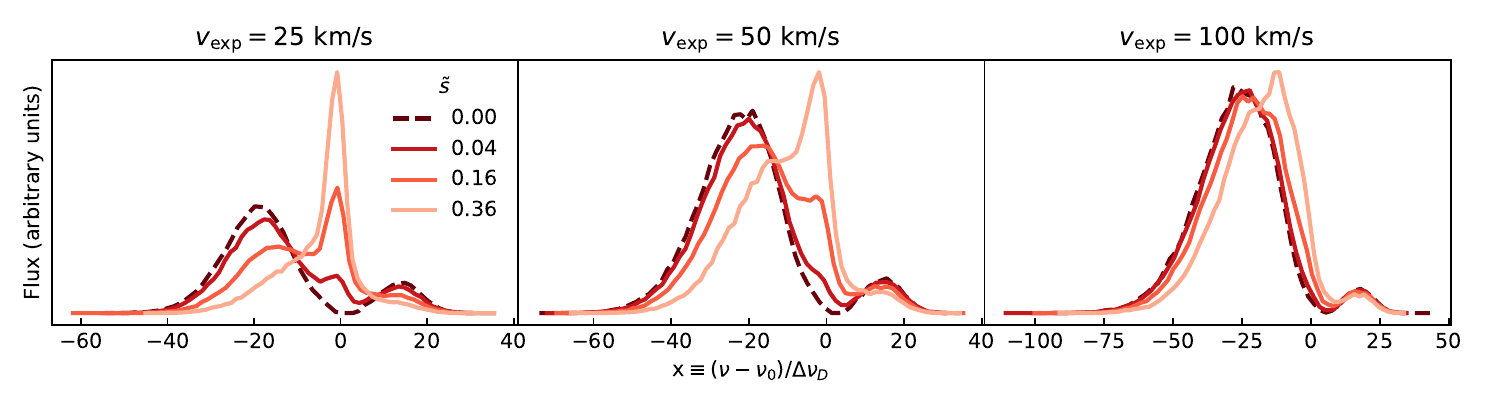}
    \caption{Emergent \Lya spectra for different hole sizes $\tilde s$ in an outflowing medium. Each panel shows the resulting line profile for a given outflow velocity $v_{\rm exp}$. For comparison, the isotropic case without a hole is shown as a black dashed line.}
    \label{fig:vel_spec}
\end{figure*}

Outflows are a common feature in star-forming galaxies, driven by various feedback processes \citep{Veilleux2005}. These bulk motions can significantly alter the radiative transfer of \Lya photons, shaping the emergent spectra in ways that differ from static media \citep[e.g.][]{bonilha_monte_1979}. While previous sections focused on static \HI slabs with escape channels, this section examines how the presence of bulk gas motion affects the behavior of \Lya photons. 

\begin{figure}
    \centering
    \includegraphics[width=0.9\linewidth]{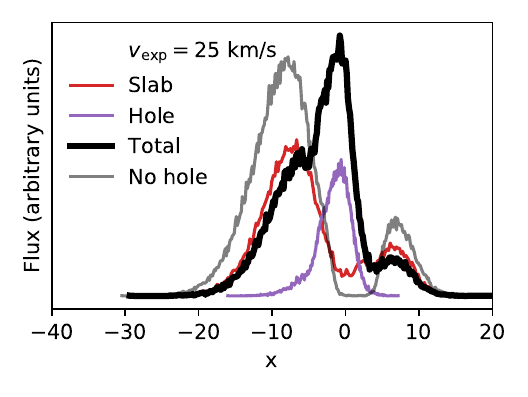}

    \caption{Emergent \Lya line profile from an outflowing medium with a hole of size $\tilde s = 0.36$, shown with a thick solid line. The spectral components corresponding to photons escaping only through the slab and only through the hole are shown with thinner lines. For comparison, the emergent spectrum from an outflowing medium without a hole is also included.}
    \label{fig:asymmetric-vel}
\end{figure}

\Lya photons propagating through a static, isotropic gas distribution undergo resonant scattering, resulting in a characteristic double-peaked emission profile, optically thick at the line center. In an outflowing medium, photons can suffer additional redshifting, facilitating their escape as the medium becomes optically thin to them \citep{ahn03,dijkstra19}.
Blueshifted photons are closer to the line center in the reference frame of the outflowing medium, where the optical depth is high, forcing some of them to undergo additional scatterings to escape towards the red side. 
This process produces an asymmetric line profile with an enhanced red peak relative to the blue peak. An example of this behavior is drawn with a dashed line in Fig.~\ref{fig:vel_spec}. 

If the outflowing medium is anisotropic, for example, featuring a low-density channel carved within the gas (a hole), the interaction of the photons escaping through it differs from that of photons propagating through denser regions. While photons emerging directly from the slab are strongly redshifted, with a suppressed blue peak, the flux escaping from the hole is also redshifted but to a lesser extent due to the (reduced) interaction with the background gas and, consequently, fewer scatterings. As a result, the hole emission lies near the line center but remains slightly redshifted. The first two panels in Fig.~\ref{fig:vel_spec} illustrate these effects in an outflowing medium with different outflow velocities and hole sizes. 

Note that when the outflow velocity is sufficiently large (e.g., third panel of Fig. \ref{fig:vel_spec}), the central peak merges with the red peak, leading to (a) the apparent disappearance of the central peak, which can misleadingly suggest the absence of a hole—even when the hole is as large as $\tilde s = 0.36$; (b) the merged central and red peaks can mimic an intrinsically less redshifted red peak; and (c) the red peak may exhibit asymmetries as a result of this blending.

Figure \ref{fig:asymmetric-vel} shows an example case with a hole $\tilde s =0.36$ at $v_{\rm exp}=25\kms$. In this figure, the slab and hole components are also shown separately, and for comparison, the no-hole case is shown, too. The resulting total spectrum may resemble a double-peaked profile, with the red peak shifted closer to the line center. This can lead to underestimating the actual outflow velocity and mask any direct evidence of a hole (i.e., no apparent central flux). Asymmetric single- or double-peaked profiles may, in fact, indicate the presence of low-density channels or holes in an outflowing medium.

The frequency shift experienced by \Lya photons during scattering arises from two effects: (i) random frequency diffusion due to the thermal motion of hydrogen atoms, and (ii) systematic Doppler shifts from the bulk velocity of the medium relative to the observer. 
A single reflection in an outflowing slab, therefore, involves both contributions. After such a reflection, the photon frequency in the observer’s frame can be written as
\begin{equation}
x_1 = x_0 - 2 \xi u
\end{equation}
where $x_0$ is the incident frequency, $u = v_{\rm exp}/v_{\rm th}$ is the outflow velocity normalized by the thermal velocity $v_{\mathrm{th}}$, and $0 < \xi \leq 1$ accounts for the angular dependence of the scattering (see, e.g., \citealp{dijkstra19}). 
Note that there is a frequency shift associated to each \textit{scattering}. However, since these scatterings are random and symmetric in frequency space most of their contributions cancel on average.
By contrast, the systematic shift due to bulk \HI kinematics does not cancel but instead contributes once per reflection. 
Accordingly, after $N$ reflections—and assuming that most reflections occur close to the backward direction relative to the incident direction (i.e., $\xi \approx 1$)—the cumulative frequency shift becomes
\begin{equation}
x_N = x_0 - 2 \xi N u.
\end{equation}

The \Lya photon is likely to have escaped after shifting to a frequency sufficiently far from the line center, reaching a regime where the medium becomes optically thin, i.e., when the optical depth satisfies $\tau(x_{ N})=N_{\rm HI}\sigma_{\rm HI}(x_{N})=1$ with HI cross section $\sigma_{\rm HI}(x)$. Assuming wing scattering in the high–column-density regime, with cross-section $\sigma_{\rm HI}(x)=\rm{a_v}/\sqrt{\pi}x^2$ and an initial frequency at line center $x_0=0$, one obtains for the number of scatterings required for escape
\begin{equation}
    N_{\mathrm{outflow}}=\frac{1}{2u \xi }\left(\frac{\tau_0 \mathrm{a_v}}{\sqrt{\pi}}\right)^{1/2}.
\end{equation}

The average number of scatterings per reflection in a static slab is $N^{\mathrm{scat}}_{\mathrm{refl}}=\left(2C\tau_0/\pi\right)^{1/2}$ (cf. \S~\ref{sec:analytic_view}), which is large enough to shift the photon to an escape frequency, leading to the rather low escaping flux through the hole (cf.~Eq.~\eqref{eq:ftilde}). However, in outflowing media, the medium becomes optically thin after fewer scatterings, resulting in a transmission probability $T = \min\{(N^{\mathrm{scat}}_{\mathrm{refl}})^{-1}, N_{\mathrm{outflow}}^{-1}\}$, i.e., where the photons escape through the slab via excursion (as discussed in \S~\ref{sec:analytic_view}, \citealp{adams_escape_1972}) or due to (subsequent) reflections leading to large frequency shifts. 

This implies that one can compute the hole-to-slab flux ratio as
\begin{equation}\label{eq:ftilde_outflows}
\tilde f=\frac{N_{\mathrm{hole}}^{-1}}{N_{\mathrm{outflow}}^{-1}}=\frac{\tilde s}{4 \xi u}\left(\frac{\tau_0 a_{\mathrm{v}}}{\sqrt{\pi}}\right)^{1/2},
\end{equation}
where $N_{\mathrm{hole}}\approx 2/\bar s$ is the number of events needed to escape through the hole as before.

\begin{figure}
    \centering
    \includegraphics[width=0.9\linewidth]{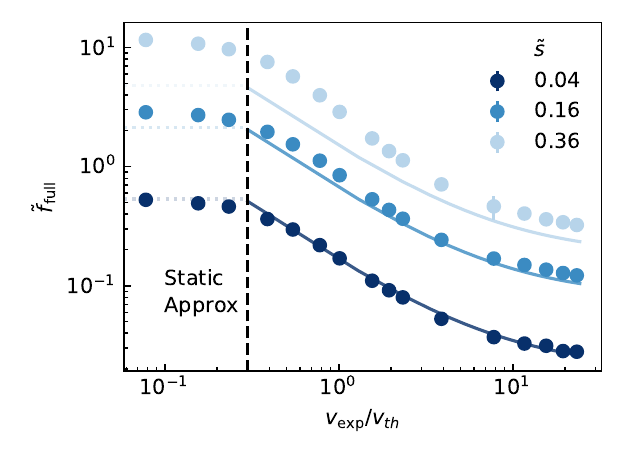}
    \caption{Flux ratio of directly escaping over scattering photons $\tilde{f}_{\rm full}$ as a function of outflow velocity for different hole sizes $\tilde s$. The analytical expectations are shown as lines: dotted segments represent the static case, while solid segments correspond to the outflow-modified regime, following Eq.~\ref{eq:ftilde_outflows} with $\xi = 0.8$. Note that here we have also included the initial escaping flux  (i.e., $\tilde f_{\rm full}\approx \tilde f +0.5 \tilde{s}$).}
    \label{fig:fvsvel}
\end{figure}

Figure~\ref{fig:fvsvel} shows the behavior of $\tilde{f}_{\rm full}$, defined as the initial escaping flux plus the usual $\tilde f$, as a function of outflow velocity $u$ for different hole sizes at a fixed column density of $N_{\mathrm{HI}} = 10^{17}\cm^{-2}$. As suggested by the figure, low outflow velocities, $u\lesssim 0.3$, $\tilde{f}_{\rm full}$ closely follow the trend observed in the static case, suggesting that the outflow has little effect in this range. Beyond this transition point, the influence of the outflow becomes significant, and $\tilde{f}_{\rm full}$ begins to decrease in agreement with Eq.~\eqref{eq:ftilde_outflows}.

\subsection{Dust}
\label{sec:dust}
Interstellar dust, produced by stellar processes and distributed throughout the ISM, plays a crucial role in shaping the propagation of \Lya photons \citep[see, e.g.][]{laursen_ly_2009}. Dust can form in evolved stars, supernova ejecta, or be transported into low-density channels by outflows or turbulence, making it plausible that even relatively clear escape paths, such as holes in the ISM, might contain some level of dust \citep{draine03}. Understanding how \Lya photons propagate in dusty, porous environments is important for interpreting observed line profiles, especially in systems where feedback processes carve out holes or channels in the ISM.

In this section, we investigate the impact of dust on \Lya escape in systems with a hole. To isolate the role of dust, we explore two scenarios: one in which dust is present only in the slab (`Dust-Free Hole'), and another where dust is distributed throughout the entire system, including the hole (`Dust-Filled Hole').

\begin{figure*}
    \centering
    \includegraphics[width=0.8
    \linewidth]{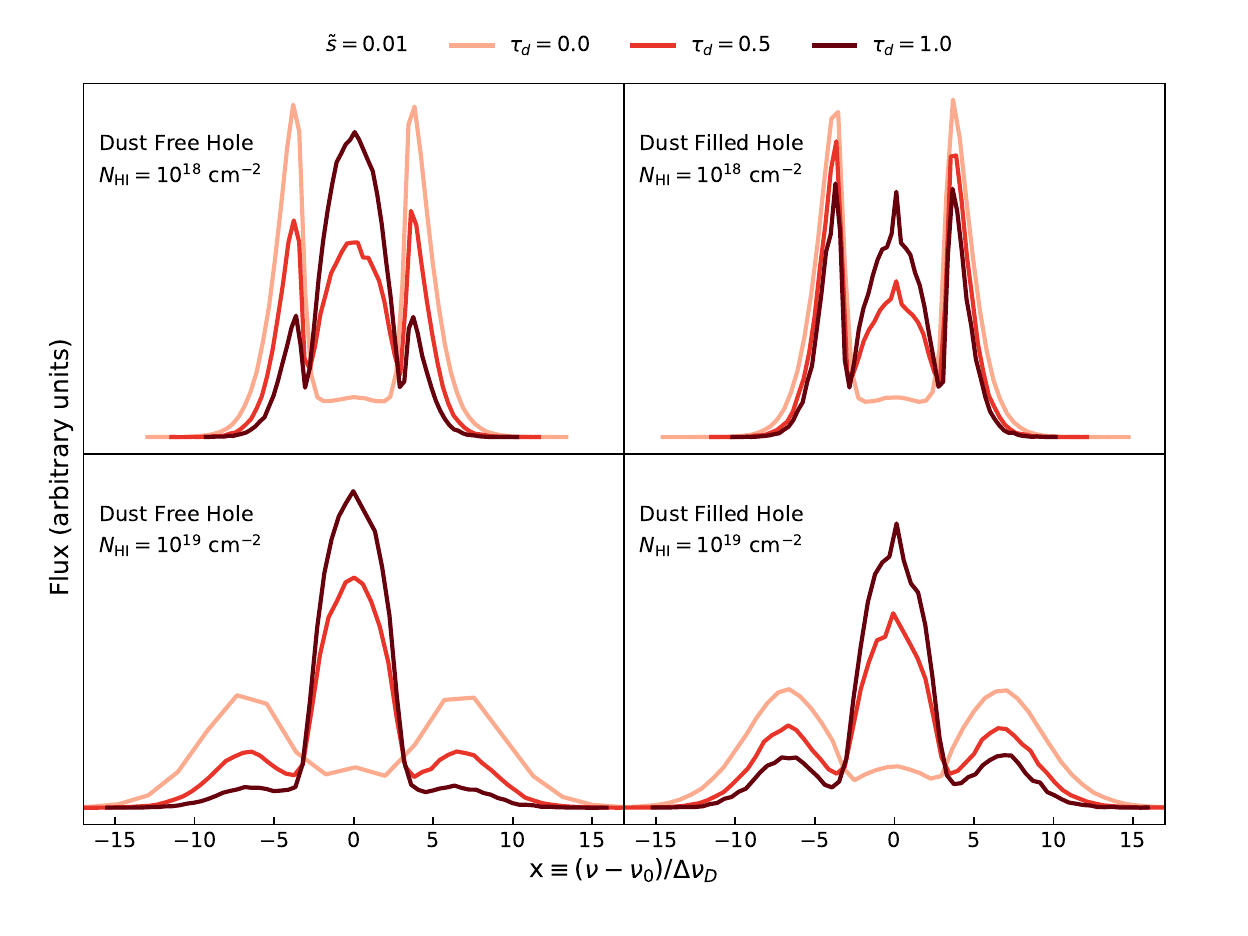}
    \caption{Emergent \Lya spectra from different configurations with dust. The spectra are normalized to unity in order to compare their profiles independently of their absolute strength. The left-hand panels correspond to models where the hole is completely free of dust, while the right-hand panels show cases where the dust content inside the hole matches that of the surrounding slab (characterized by optical depth $\tau_\mathrm{d}$)} 
    \label{fig:dustspec}
\end{figure*}

Figure~\ref{fig:dustspec} shows the emerging \Lya spectra from a slab with a hole of size $\tilde s = 0.01$, for two different \HI column densities and various dust optical depths $\tau_{\rm d} = 0-1$. The left-hand panels correspond to cases where the hole is dust-free, while the right-hand panels show cases where the dust content inside the hole matches that of the surrounding slab. In both configurations, the presence of dust enhances the flux near the line center, even when the hole itself contains dust. Compare, for instance, the $\tau_{\mathrm{d}} = 0$ case, where central flux is minimal, with the $\tau_{\mathrm{d}} = 1$ case, where dust contributes to higher flux around the line center.

Note in Fig.~\ref{fig:dustspec} that when both the slab and the hole are entirely free of dust ($\tau_{\mathrm{d}} = 0$), the central flux is minimal, and the hole remains effectively ``hidden'' in the spectral line. In contrast, when $\tau_{\mathrm{d}} >0$, the central flux is enhanced in relative terms, making the hole's presence more apparent. 

The fraction of \Lya photons that escape from a hole that contains dust but no gas is simply, 
\begin{equation}\label{eq:feschole}
f_{\mathrm{esc}}^{\mathrm{hole}}=\mathrm{e}^{-\bar\alpha(1-A)\tau_{\mathrm{d,{hole}}}},    
\end{equation} 
where $A$ is the albedo (see §~\ref{sec:numerical_setup} for further details on the dust model), $\tau_{\mathrm{d,{hole}}}$ is the dust optical depth in the hole, and $\bar\alpha \approx \sqrt{2}$ is a constant that accounts for the random angles at which photons traverse the hole. Note that Eq.~\eqref{eq:feschole} is also true for the dust-free case as it becomes $f_{\rm esc}^{\mathrm{hole}}=1$ with $\tau_{\mathrm{d,{hole}}}=0$.

On the other hand, photons escaping through the slab interact with both gas and dust. To account for this, we use the analytical solution derived in \cite{neufeld_transfer_1990}, 
\begin{equation}\label{eq:fescslab}
f_{\mathrm{esc}}^{\mathrm{slab}}= \frac{1}{\cosh \left( \frac{\sqrt{3}}{\pi^{5/12} \chi} \left( (a\tau_0)^{1/3} \tau_\mathrm{d,abs} \right)^{1/2} \right)},    
\end{equation}
where $\chi=0.5$ is a fitting parameter and $\tau_{\mathrm{abs}}=(1-A_\mathrm{d})\tau_{\mathrm{d}}$ is the dust absorption optical depth. 

\begin{figure}
    \centering
    \includegraphics[width=0.9\linewidth]{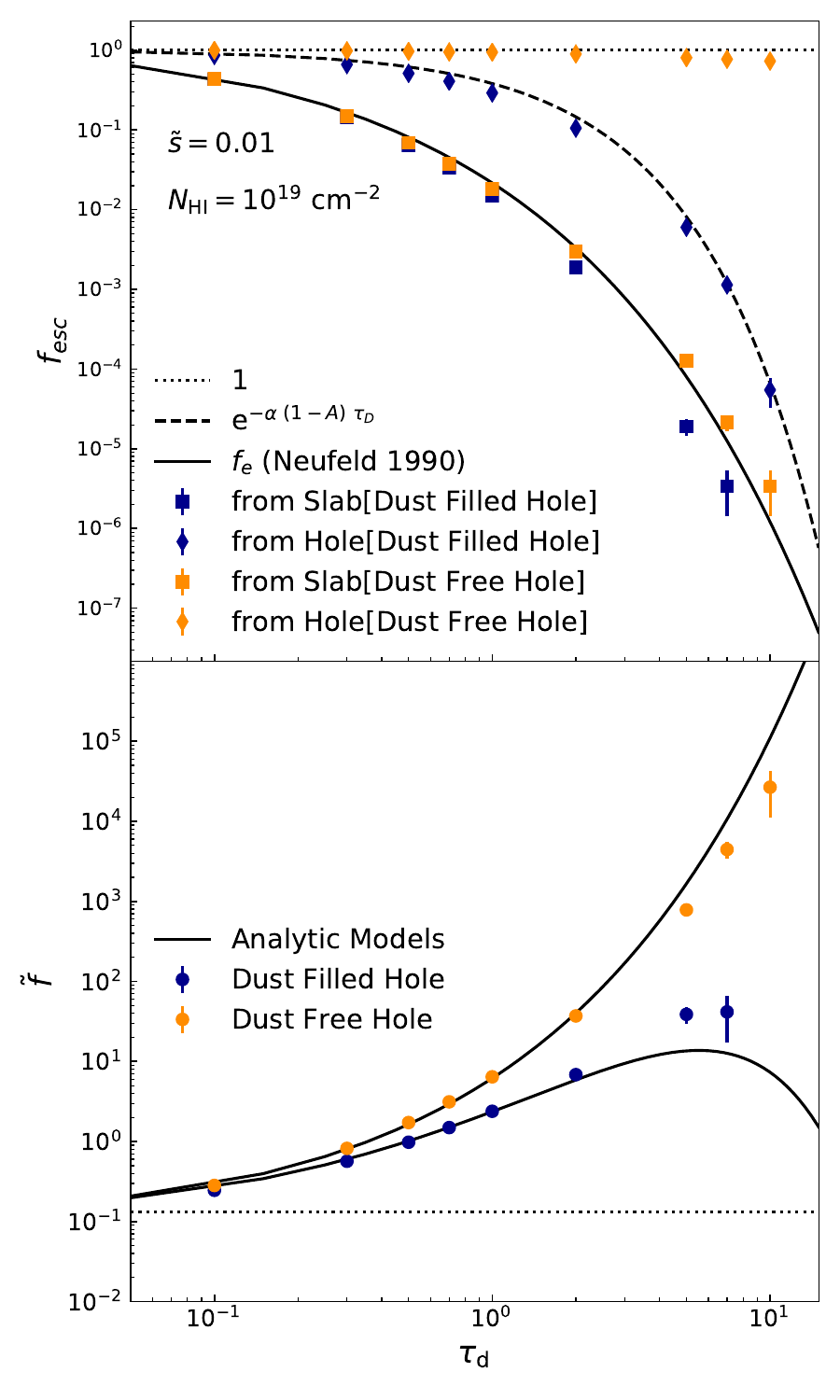}
    \caption{Effect of a dusty medium on \Lya escape and line profile. \textbf{Upper panel:} \Lya escape fractions for the individual components. We show the escape fractions from the slab and the hole separately for both the Dust-free hole and the Dust-filled hole configurations. Black lines indicate the expected analytical trends as given by Eqs.~\ref{eq:feschole} and \ref{eq:fescslab}. \textbf{Lower panel:} The flux ratio $\tilde{f}$ as a function of dust optical depth for both configurations. Solid lines show the analytical predictions from Eq.~\ref{eq:ftilde_dust}.}
    \label{fig:fescapevstauD}
\end{figure}

To test the analytical predictions, we tracked photons based on whether they escaped through the slab or the hole, and evaluated their escape fractions separately. The escape fraction is defined as the ratio of photons emerging from either the dusty hole or slab to the number that would escape from the same region in a completely dust-free medium. The top panel of Fig.~\ref{fig:fescapevstauD} shows these escape fractions as a function of dust optical depth, separated into their individual components: a dusty slab, an empty hole, and a dusty hole. The analytical predictions from Eqs.~\eqref{eq:feschole} and \eqref{eq:fescslab} are shown and match the simulation results well.

Finally, we compute the hole-to-slab flux ratio, $\tilde{f}$, to evaluate how dust alters the relative contribution of photons to the \Lya line profile in systems with an escape channel. Since each component contributing to $\tilde{f}$ experiences extinction depending on its dust content, the ratio is expected to vary by their respective escape fractions. This can be expressed as:
\begin{equation}\label{eq:ftilde_dust}
\tilde{f}_{\rm d}=\tilde{f}\ \frac{f_{\rm esc}^{\mathrm{hole}}}{f_{\rm esc}^{\rm slab}}    
\end{equation}
where $\tilde{f}$ corresponds to the case where dust is completely absent (i.e. given by Eq.~\eqref{eq:ftilde}).

The lower panel of Fig.~\ref{fig:fescapevstauD} shows $\tilde f_{\rm d}$ as a function of the dust content $\tau_{\mathrm{d}}$. Consistent with what we observed in Figure~\ref{fig:dustspec}, dust increases the relative flux emerging near the line center by promoting photon escape through the hole. This enhancement is strongest when the hole is dust-free, while in dusty holes the effect is more subdued due to absorption and scattering that reduce the probability of photons escaping directly through the hole. As the dust optical depth increases,$\tilde f_d$ also grows, further supporting the idea that dust could reveal the presence of a hole in the spectrum by enabling the emergence of central flux that would otherwise be absent in a dust-free medium.

\subsection{Lognormal distribution}
\label{sec:lognormal}
While in the above, we considered media with constant number densities, in reality, the neutral hydrogen distributions are more complex. Specifically, the interstellar medium (ISM) and the HI around star-forming regions are shaped by feedback and turbulent processes, which naturally lead to lognormal density distributions. Such distributions have been shown to arise from supersonic turbulence in the ISM, with their variance closely tied to the turbulent Mach number and the nature of the driving forces \citep{federrath_comparing_2010,Klessen2014}.
As a step toward capturing some of this complexity, we constructed simplified slab configurations (Figure~\ref{fig:diagram}) by assigning a single column density to each cell within the slab (i.e., with depth $d$ and $100\times 100$ values are drawn; cf. \S~\ref{sec:geometry}) from a lognormal distribution. It is common to characterize lognormal distributions by two parameters $(\mu,\,\sigma)$ where the former controls predominantly the centroid, and the latter the width of the distribution.

We performed separate simulations in which the mean ($\exp(\mu+\sigma^2/2)$), median ($\exp(\mu)$), or mode ($\exp(\mu-\sigma^2$) of the lognormal distributions was fixed to $N_{\mathrm{HI}} = 10^{19}$ or $10^{20}\cm^{-2}$. We then explored how \Lya photons interact with such inhomogeneous media, focusing on which column densities are effectively probed by \Lya photons.

\begin{figure*}
    \centering
    \includegraphics[width=0.97\linewidth]{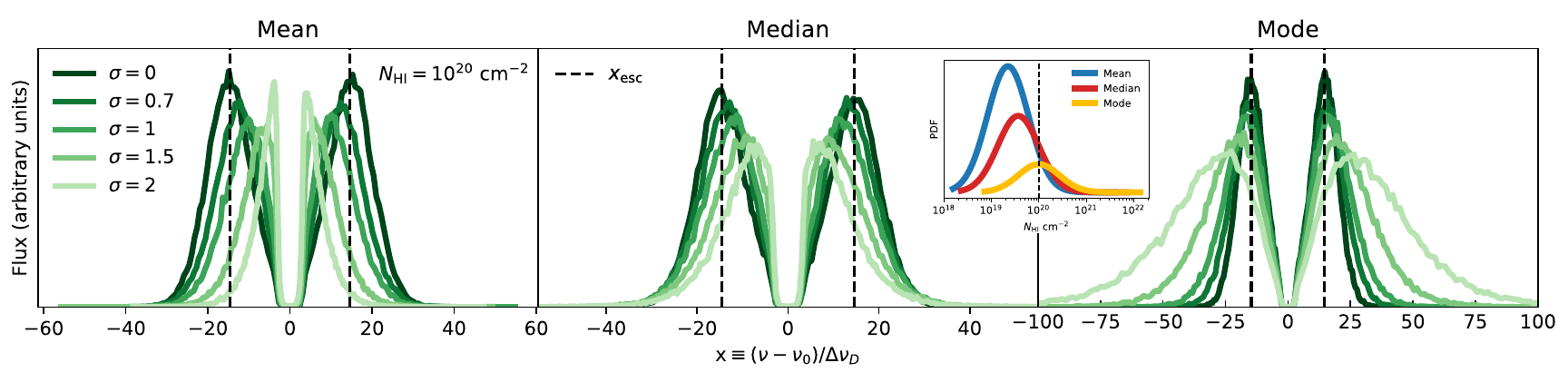}
    \caption{Emergent \Lya spectra for lognormal column density distributions with increasing dispersion $\sigma$. Each panel shows a separate case where either the mean (left), median (middle), or mode (right) is fixed to $N_{\mathrm{HI}} = 10^{20}\cm^{-2}$. Curves within each panel correspond to different values of $\sigma$, with lighter lines corresponding to larger values of $\sigma$, and thus greater inhomogeneity in the column density distribution. The darkest curve represents the homogeneous case ($\sigma = 0$). Dashed vertical lines mark the expected peak positions for a homogeneous slab with the same column density. The inset shows a comparison of the lognormal PDFs.}
    \label{fig:spectra_lognorm}
\end{figure*}

Figure~\ref{fig:spectra_lognorm} shows the emergent \Lya spectra for grids with increasing $\sigma$. All panels refer to the same reference column density, but each panel represents a different case in which either the mean (left), median (central), or mode (right panel) of the distribution is fixed to the value of $N_{\rm HI}=10^{20}\cm^{-2}$. Within each panel, curves of varying color show the impact of increasing inhomogeneity: lighter colors correspond to larger $\sigma$, while the darkest curve represents the homogeneous limit ($\sigma = 0$). The dashed vertical lines point the expected peak positions for a homogeneous slab with the same column density.

The first panel of Fig.~\ref{fig:spectra_lognorm} (in which the mean of the lognormals was fixed) shows that for increasing $\sigma$, the emergent spectra become narrower (the peak separation decreases) more skewed, i.e., they exhibit extended wings. These trends imply that \Lya photons are susceptible to the distribution of column densities, in particular, the width of the column density distribution matters (and the spectrum is not simply set by the mean). The narrowing of the spectra with increasing $\sigma$ also implies that \Lya photons are more likely to escape (and thus probe) lower column densities; although the wings indicate that a non-negligible fraction of photons also escapes through pathways of high optical depth (see  Appendix~\ref{sec:appendix_NHI_probed} where we analyze the column density distribution traversed by the photons).

A similar, albeit weaker, behavior can be observed in the central panel of Fig.~\ref{fig:spectra_lognorm} where we fixed the median of the distributions to $N_{\rm HI}=10^{20}\cm^{-2}$. Here, again, the peak separation decreases with increasing $\sigma$, while the outer wings stay approximately constant, effectively increasing the skewness of the peaks. However, it is clear that the emergent \Lya spectrum is better defined through the median of the distribution compared to the mean, which is in line with our above discussed findings (where $\tilde f\sim \tilde s$).

The behavior is different in the third panel of Fig.~\ref{fig:spectra_lognorm}, where the mode is fixed. Here, the fixed column density corresponds to the most probable value encountered by the photons. As $\sigma$ increases, the distribution becomes broader, and the relative difference in probability between the mode and the high-column-density tail decreases. In other words, high-$N_{\rm HI}$ regions become more comparable in weight to the peak of the distribution. Consequently, \Lya photons are more likely to interact with these higher-density regions as $\sigma$ increases, resulting in an increase in peak separation (the opposite trend from the mean and median cases). Moreover, the spectra in this case become visibly broader, with thicker wings compared to the mean and median cases. This is a clear signature of enhanced interaction with the high-$N_{\rm HI}$ tail of the distribution. 
This shows that contrary to the common assumption that \Lya photons escape only through low column density gas (the path of the least resistance), they probe a broad range of values across the distribution.

\begin{figure}
    \centering
    \includegraphics[width=0.9\linewidth]{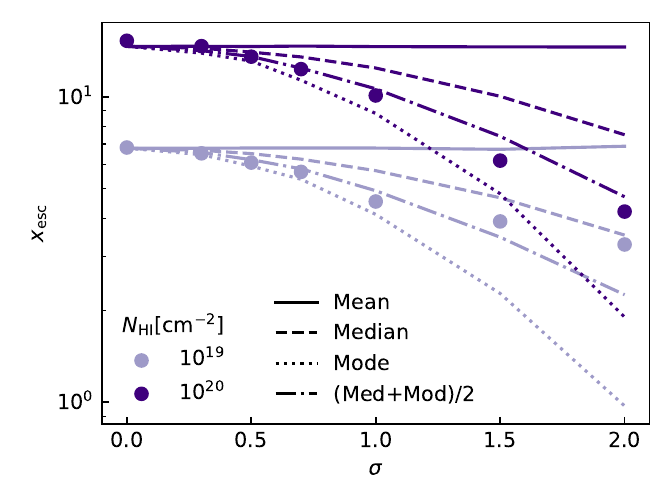}
    \caption{Comparison of theoretical and simulated \Lya peak escape frequencies, $x_{\rm esc}$, for slabs with lognormal column density distributions of varying dispersion $\sigma$ and fixed mean. Theoretical predictions (lines) are based on the mean, median, and mode column densities of each distribution. The dot-dashed line shows an intermediate value between the mode and median predictions. The simulated peak escape frequencies (black points) fall between the mode and median estimates rather than aligning with the lowest column density (“path of least resistance”), demonstrating that escaping \Lya photons probe a range of column densities rather than only the lowest-density regions.}
    \label{fig:x_esc}
\end{figure}

While a range of column densities is probed by \Lya photons, it is clear from Fig.~\ref{fig:spectra_lognorm} that they seem closer in probing the median than the mean.
To explore this quantitatively, we show in Fig.~\ref{fig:x_esc} the peak position $x_{\rm esc}$ of the emergent spectra, and compare it to the values expected from a homogeneous slab ($x_{\rm esc}=1.1 \left(\mathrm{a_v} \tau_0/\sqrt{\pi}\right)^{1/3}$; \citealp{adams_escape_1972}). In the simulations the mean was fixed to $10^{19}\,\cm^{-2}$ and $10^{20}\,\cm^{-2}$ and $\sigma$ was varied.
Fig.~\ref{fig:x_esc} shows how the analytical $x_{\rm esc}$ changes with $\sigma$ if a corresponding mean or mode is considered.
As discussed above, the peaks are much further apart than in a slab with the corresponding mean and slightly further apart than a slab with the same median, but significantly closer together than in a slab with the same mode of the distribution. 

\begin{figure}
    \centering
    \includegraphics[width=0.95\linewidth]{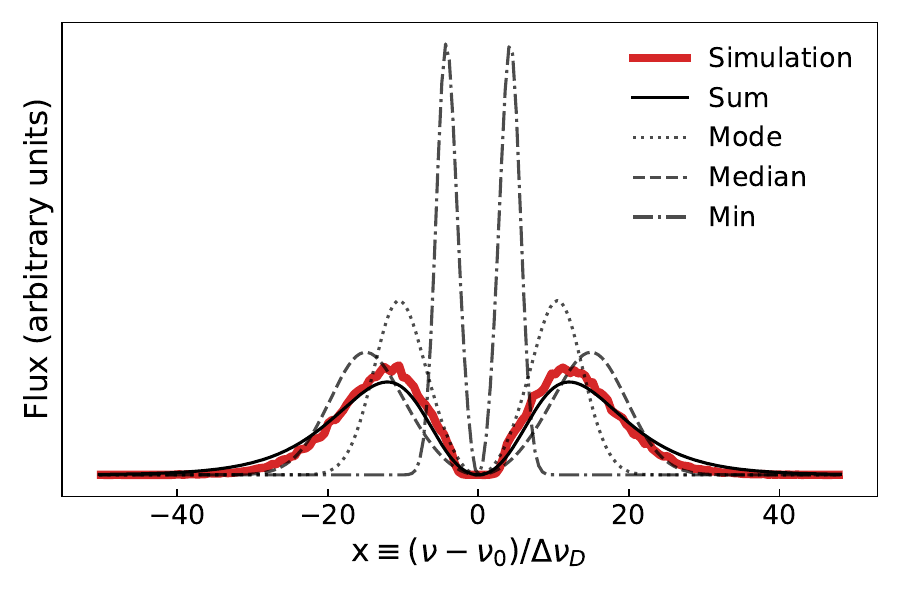}
    \caption{Comparison between the emergent \Lya spectra from a slab with a lognormal column density distribution obtained via full radiative transfer simulations (thicker solid line) and the weighted sum of analytical Neufeld line profiles (thinner solid line), where the weights are given by the lognormal probability density. Both profiles show close agreement. Overplotted are Neufeld profiles calculated for the minimum, median, and mode column densities of the distribution. The emergent \Lya line profile does not trace the lowest column density in the distribution, but instead corresponds to a value between the median and the mode.}\label{fig:sum_log}
\end{figure}

In Section~\ref{sec:filled_hole}, we show that the emergent \Lya spectrum from a slab composed of two distinct column density regions (a uniform slab with a  lower-density hole) can simultaneously reflect both components. In that case, the resulting spectrum could be well approximated by a superposition of two individual \Lya line profiles, each corresponding to one of the two column densities, summed according to the density distribution (two delta functions in this case). 

In Figure~\ref{fig:sum_log}, we explore whether this additive approach can also reproduce the emergent spectra from a continuous, lognormal distribution of column densities. For this purpose, we compute analytical \Lya line profiles using the \citet{neufeld_transfer_1990} analytical solution for a range of column density values spanning the lognormal distribution. Each profile is then weighted by the normalized lognormal probability density and summed to produce a composite spectrum (black solid line). This spectrum is compared to the full radiative transfer result obtained from the inhomogeneous lognormal grid (red solid line). The two profiles agree closely, both in peak positions and overall line shape.

\section{Discussion}
\label{sec:discussion}

\subsection{What gas do \texorpdfstring{\Lya}{Lya} photons probe?}
\label{sec:what_lya_probes}
Being the most prominent resonant line in the Universe, \Lya observables are commonly used to constrain cold ($\sim 10^4\,$K) gas distributions in the ISM and CGM of galaxies.  In particular, the wide variety of \Lya spectra (see, e.g., the `Lyman Alpha Spectral Database' \citealp{runnholm_lyman_2021}\footnote{\url{http://lasd.lyman-alpha.com}.}) can reveal information about bulk velocities and neutral hydrogen column densities. A common way to extract such information is through fitting observed spectra using simplified models such a the `shell model' \citep{ahn_ly_2002,verhamme_3d_2006,gronke_modeling_2017} or clumpy, multiphase models \citep{2021MNRAS.502.2389L,Li2022}. These models can not only produce the complex observed spectra well but also yield estimates and uncertainties on key parameters such as the HI column density, bulk motions, temperatures and turbulent motion of the scattering medium. 
However, it is unclear what this scattering medium exactly is -- and in turn how one can interpret the properties obtained for \Lya spectra modeling.

Is, for instance, the HI column density $N_{\rm HI}$ obtained the mean, line-of-sight or minimum column density of a given galaxy? Many arguments speak for the latter, i.e., that \Lya photons are susceptible to the `path of least resistance'. This was, for instance, argued in \citet{Vielfaure2020A&A...641A..30V,Vielfaure2021A&A...653A..83V} where the authors could measure the line-of-sight HI column density through the afterglow of gamma-ray bursts (GRBs), or \citet{Herenz2016A&A...587A..78H, herenz2025} where the authors tried to correlate regions of enhanced turbulence with \Lya escape. 

This scenario would imply that the parameters measured in this way correspond to pathways of low optical depth, that is, in a more realistic column density (and kinematic property) distribution, they are heavily skewed towards low $N_{\rm HI}$, and high (bulk/turbulent) velocities. 
This would have implications for, e.g., the tool to detect the escape of hydrogen ionizing photons (which are thought to `leak' out of these low $N_{\HI}\lesssim 10^{17}\,\cm^{-2}$ channels). We will discuss this point more in \S~\ref{sec:LyaLyC}.

Our results show clearly that instead \Lya photons probe more global properties of galaxies. In fact, as the results in \S~\ref{sec:lognormal} show, the emergent spectrum corresponds closest to the median HI column density. This susceptibility to median properties holds great potential as one can use \Lya spectral modeling to constrain galaxy properties rather than extrema statistics. 

For instance, using the `shell model' fitted $N_{\rm HI}$, one can infer the total HI mass of galaxies, median in-/outflows (and corresponding mass fluxes), and overall levels of turbulence. 
Naturally, in order to test this hypothesis, one needs to correlate these quantities obtained from \Lya fitting with auxiliary data, such as that obtained from 21cm.

\subsection{\texorpdfstring{\Lya}{Lya} as a LyC tracer: where are they escaping from?}
\label{sec:LyaLyC}
Arguably, the biggest puzzle regarding the Epoch of Reionization (EoR) is how ionizing radiation escapes from the thick neutral hydrogen (HI) around galaxies into the intergalactic medium \citep[see reviews by][]{barkana_beginning_2001,klessen_first_2023}. Because \Lya photons resonantly scatter in the same HI that absorbs Lyman continuum (LyC) photons, and are at the same time much easier to observe, \Lya can be regarded as a prime proxy for LyC escape \citep[e.g.][]{verhamme_using_2015,dijkstra_ly-lyc_2016}. 

Understanding this connection requires examining the conditions under which LyC can actually escape from galaxies. Young star-forming regions are usually surrounded by dense gas, which makes it difficult for LyC photons to escape. Only after massive stars die can feedback mechanisms clear out gas, creating low-column-density channels. It is through these channels that LyC photons can finally escape. These same channels should also affect \Lya emission, since \Lya is highly sensitive to the gas distribution. Thus, studying how \Lya photons propagate through these structures offers a way to infer the conditions that regulate LyC escape. Earlier work has shown that in some cases \Lya escape can be highly anisotropic in such geometries \citep{behrens_inclination_2014, zheng_anisotropic_2014}. However, we still lack a systematic picture of what fraction of \Lya photons actually escape through these same channels. This gap limits our ability to use \Lya as a reliable proxy for LyC leakage.

Observationally, a direct connection between \Lya observables and escape fraction of ionizing photons is only possible through low-redshift galaxies where LyC photons are still detectable. These measurements can then be used to constrain predictions for galaxies at higher redshifts. Works that study these low-redshift samples report an anti-correlation between \Lya peak separation ($v_{\rm sep}$) and LyC escape fraction ($f_{\rm esc,LyC}$) \citep{izotov_detection_2016,izotov_low-redshift_2018,izotov_lyman_2021,flury_low-redshift_2022-1,flury_low-redshift_2022}. In these galaxies, smaller $v_{\rm sep}$ values correspond to fewer scattering events, which means that photons traverse lower-column-density regions. Most confirmed LyC leakers in these works show relatively narrow peak separations (often $v_{\rm sep} \lesssim 300$ km/s)

To expand to higher redshift samples, \citet{kerutt_lyman_2024} presented a large LyC leaker search in $\sim2000$ VLT MUSE \Lya emitters (LAEs) at $z=3-4.5$. They report high escape fractions but find no correlation with \Lya line properties. In particular, they find that galaxies in their sample with peak separations $v_{\rm sep}\gtrsim 400 \rm{km/s}$ and escape fractions $f_{\rm esc,LyC}=22-88$\%, which is inconsistent with the anti-correlation found for lower redshift samples. \citet{pahl_ly_2024} also found this inconsistency, measuring the offset between the red-peak and the systemic velocity. Furthermore, \citet{tang_ly_2024} found that small \Lya peak separations ($\lesssim 100$ km/s) are rare at $z \sim 5-6$, making $v_{\rm sep}$. 

These findings support the picture proposed in this work, i.e., that \Lya photons do not probe only the `extreme statistics', that is, the low column density channels through which LyC photons can escape but rather a global HI column density. This would have several implications for the usage of \Lya to trace ionizing photon escape:
\begin{enumerate}
    \item there should be many `hidden leakers' (such as the ones found by \citealp{kerutt_lyman_2024}) out there in which the measures (LOS) escape fraction $f_{\rm esc,LyC}>0$ but the \Lya observables predict an overall large $\gg 10^{17}\cm^{-2}$ HI column density.
    \item specifically, this implies that even for galaxies where $f_{\rm esc,LyC}\approx 0$ is measured, and, e.g., the \Lya peak separation is large, the global LyC escape can be non-negligible,
    \item in the known LyC leakers in which $f_{\rm esc,LyC}$ agrees with the \Lya prediction (such as the low-$z$ samples discussed above) HI gas is most likely distributed fairly isotroptically, thus, the LOS escape fraction agrees with the `global' one.
\end{enumerate} 
Testing this hypothesis is non-trivial, as observations are constrained to the LOS ionizing escape fractions. However, strategies searching for LyC leakage in a `blind' fashion in large samples of LAEs \citep[such as][]{kerutt_lyman_2024} can reveal in which cases the \Lya observables do (not) agree. 
Another possibility to disentangle between the LOS and global escape fractions is to use ionization structures in different directions \citep[as, e.g., done in][]{Herenz2023} or to infer it energetically \citep[e.g.][]{Barger2013,Niederhofer2016} -- however, both are challenging at higher redshifts.

While theoretical, numerical, and observational studies are needed in order to pin down the connection between \Lya and LyC escape, we clearly show in this study that the simple picture of \Lya and LyC escaping through the same pathways is not correct. Instead, \Lya gives us a more comprehensive view of the HI distribution in and around galaxies (cf. \S~\ref{sec:what_lya_probes}), and their connection can help in constraining the global ionizing budget a galaxy contributes to the IGM.

\subsection{Where are all the triple peaks?}
The \Lya emission line is a powerful probe of the interstellar and circumgalactic medium (ISM and CGM) of galaxies across cosmic time. Radiative transfer through anisotropic gas can produce a wide variety of \Lya line morphologies, with single- and double-peaked profiles being by far the most commonly observed \citep[e.g.,][]{erb_ly_2014, yang_green_2016, runnholm_lyman_2021}. Double-peaked profiles, featuring a blue and a red peak, result from resonant scattering in optically thick gas, where outflows and inflows shape the strength and separation of the peaks. In many cases, however, the blue peak is suppressed due to galactic outflows and IGM absorption (particularly at high redshift), leaving only a single redshifted peak \citep[e.g.][]{laursen_intergalactic_2011,byrohl_variations_2020}.

In a small subset of spectra, however, an additional emission component emerges near the systemic velocity, giving rise to the distinctive and still uncommon triple-peaked \Lya profile. The central peak forms due to \Lya photons directly escaping through low-column-density channels \citep{zackrisson_spectral_2013, behrens_inclination_2014, behrens_beamed_2014}, which were likely carved by feedback processes in the ISM. The reduced scattering within these channels allows photons to remain close to line center. While such profiles have long been considered rare, with only a few reported cases \citep[e.g.,][]{vanzella_direct_2018, rivera-thorsen_sunburst_2017,izotov_low-redshift_2018}, more recent studies have begun to identify them in larger samples. For example, \citet{vitte_muse_2025} report that about 5\% of galaxies in a sample from the MUSE extremely Deep Field show triple-peaked spectra, and in the past few years, works such as \citet{tang_ly_2024} and \citep{Vanzella2020MNRAS.491.1093V} have noted individual galaxies with triple-peaked \Lya emission.\footnote{Note, however, the potential misidentification of \Lya peaks which has been discussed, e.g., by \citet{protusova_unique_2024} who studied a rare double-peaked profile at $z\approx 6.6$ and noted that it could in fact be a triple peak with a very faint blue component.}

This rarity is surprising, given that we know from theoretical \citep[e.g.,][]{mckee_theory_1977, clarke_galactic_2002}, observational \citep[e.g.,][]{bagetakos_fine-scale_2010, heckman_extreme_2011, alexandroff_indirect_2015} and simulation studies \citep[e.g.,][]{avillez_global_2005, walch_silcc_2015, kim_three-phase_2017} that gas distributions are highly anisotropic and porous, a geometry that should favor the escape of \Lya photons through both optically thick and low-column-density channels. The unexpectedly low fraction of triple-peaked detections might suggest that such channels are either rare (though the evidence suggests otherwise) and/or short-lived, or that they are common in the gas distribution but often hidden by a combination of observational limitations, classification biases, or intrinsic gas properties that can suppress or misidentify the central peak.

As we show in this work, however, \Lya photons do not exclusively follow paths of least resistance but can also traverse regions of high column density. Suppose low-column-density channels occupy only a small area fraction of the sight lines (i.e., the channel is small). In that case, the resulting central peak may be faint relative to the red and blue peaks (in agreement with Eq~\ref{eq:ftilde}), either requiring higher spectral resolution to detect, or being lost amid the stronger red and blue peaks. Other gas properties also influence the relative flux $\tilde{f}$ through channels and optically thick gas. For example, inflows, outflows, or turbulence can shift or broaden the peaks, which could cause the central component to blend with the sides (see \S\ref{sec:outflows} for discussion). Filled channels with neutral gas can also cause partial merging of peaks (see \S\ref{sec:filled_hole}), leaving only subtle asymmetries that may be overlooked in observations. Dust can also modify the profile by enhancing the relative strength of the central peak (see \S\ref{sec:dust}). This could result in a single-peaked profile that might be misidentified as a single peak or a double peak with the apparent absence of the blue peak due to circumgalactic absorption. \\

In summary, our results show that the rarity of triple-peaked \Lya profiles may not reflect a true lack of low-column-density channels, but are rather due to the complex interplay of gas geometry, kinematics, and resonant line radiative transfer. 
While revealing the further triple-peaked \Lya emission requires high-resolution, high-sensitivity spectroscopy, uncovering them observationally -- combined with insights from radiative transfer -- can give valuable insights  into the physical conditions and distribution of the gas of these galaxies, and offers a unique window into the structure, dynamics, and feedback processes shaping the ISM and CGM of galaxies across cosmic time.

\section{Conclusions}
\label{sec:conclusions}
While \Lya plays a pivotal role in astrophysics, the theory of anisotropic \Lya escape is still underdeveloped.
In this work, we investigated the escape of \Lya radiation from porous neutral gas distributions using Monte Carlo radiative transfer simulations. 

Our main conclusions are as follows:
\begin{enumerate}
    \item \Lya are less prone to escape through low-density pathways and instead have a (maybe surprising) large probability to escape through pathways of high optical depths \citepalias[in agreement with][]{almadamonter_crossing_2024}.
    This has the interesting implication that when considering deeper `hallway' like channels (as opposed to thinner `door' like) \Lya photons penetrate the side-walls of the channel leading to a significantly reduced flux at line center -- and do not simply scatter off them leading to a `beamed' \Lya escape as previously thought. 
    
    \item When subdividing the low-$\tau$ channel into several, we found that this has little impact on the emergent \Lya spectrum or total escaping flux. 
    This suggests that the level of porosity is less significant for \Lya escape than their overall geometry and covering fraction, meaning \Lya spectra can provide direct constraints on the size and shape of the pathways through which photons escape.

\item Non-empty channels, e.g., channels containing a fiducial residual neutral fraction, modify \Lya escape and the emergent spectra. When the neutral hydrogen inside the channels is optically thin for \Lya photons ($N_{\rm HI} \lesssim 10^{13},\mathrm{cm^{-2}}$), they escape in a similar way to the empty hole case. At higher densities within the hole, scattering reduces the central flux and introduces additional spectral features such as `quadruple peaks' if the density contrasts between the channel and the surrounding medium are sufficiently large. We also show that in this case the emergent spectrum can be reconstructed using our derived hole-to-flux ratio $\tilde f$.

    \item Outflows play a key role in shaping \Lya escape. By redshifting photons, outflowing gas reduces resonant scattering and produces asymmetric line profiles. Low-column-density channels within these outflows can generate a central, moderately redshifted peak. However, this peak often blends with the dominant red peak, leading to a more skewed or multi-peaked spectrum. This can make ionized channels difficult to detect. We developed an analytical framework that can predict the flux ratio escaping through the presence, as well as the emergent spectra, also in the presence of outflows. This can be used in the future to constrain the extent of low-$N_{\rm HI}$ pathways.
    
    \item Dust in the ISM modulates \Lya escape, even in empty channels. We compare two idealized cases: dust-free holes and dust-filled holes (mimicking channels carved, e.g., through supernovae and radiative feedback, respectively). 
    Compared to the case without dust, dust-free holes generally enhance flux near the line center, making small channels that would remain hidden in a dust-free medium more visible in the spectra. Absorption and scattering reduce this effect in dust-filled holes. In some cases, only the central peak may be observed. Using previous predictions for $f_{\rm esc}$ in a homogeneous medium, we constructed a model that can predict the overall \Lya escape as well as the emergent spectrum in a dusty, anisotropic medium.
    
    \item Moving away from the simple `channel in homogeneous slab' setup, we also studied \Lya escape through more realistic inhomogeneous media with lognormally distributed column densities.
    We confirmed that, also in this case, \Lya photons do not escape solely along the lowest-density paths but instead through a broad range of column densities. This behavior is reflected in the emergent \Lya spectra. Specifically, we found that the spectra peak positions are located between the corresponding homogeneous slab of the median and mode of the distribution, confirming that \Lya photons probe the overall structure of the medium rather than just the ``path of least resistance''. We also show that the (more skewed and broader) spectrum of the lognormal distribution can be approximated by a simple weighted sum of the corresponding homogeneous spectra.

\item We discuss our findings in light of observations and their implications for the escape of ionizing photons.
Specifically, we speculate that \Lya spectra suggesting high column densities may still harbor narrow, low-covering-fraction channels through which LyC photons can escape, and in rare cases, direct alignment with such a channel would produce very high LOS LyC escape despite an otherwise opaque spectrum. Conversely, a \Lya spectrum indicating low column density ($N_{\rm HI} \lesssim 10^{17},{\rm cm}^{-2}$) generally points to a high global LyC escape fraction, rather than just a single favorable sightline.
\end{enumerate}
While further work is required to fully bridge the gap to realistic galaxy environments, this study significantly extends our previous efforts by incorporating dust, outflows, and complex geometries. In doing so, it clarifies how the presence of low-density channels and anisotropic gas structures alters emergent \Lya line profiles and identifies the spectral features that reflect specific underlying physical conditions. Specifically, this work demonstrates that \Lya emission does not preferentially trace the lowest column-density channels but instead reflects more general properties of the neutral gas.

\section*{Acknowledgments}
We thank David Neufeld for the very useful discussions and encouragement. SAM thanks the members of the Multiphase Gas at MPA for the discussions and support. MG thanks the Max Planck Society for support through the Max Planck Research Group, and the European Union for support through ERC-2024-STG 101165038 (ReMMU).

\section*{Data Availability}
The data underlying this paper will be shared on reasonable request to the corresponding author.



\bibliographystyle{mnras}
\bibliography{references,reference_sj} 




\section{Column density perceived by photons in the lognormal case}
\label{sec:appendix_NHI_probed}
\begin{figure}
    \centering
    \includegraphics[width=0.9\linewidth]{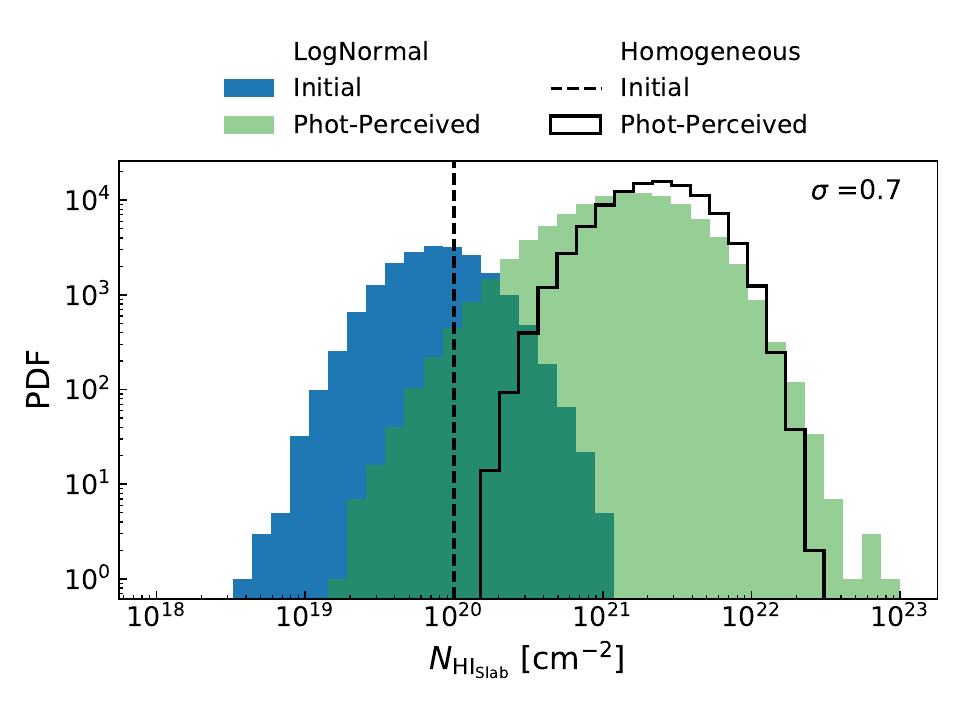}
    \caption{Column density distributions for the initial and for the photon-perceived distributions. The photon-perceived distribution of $N_{\rm HI}$ in the homogeneous slab is shown  as an outline, the initial delta function for the homogeneous distribution as a vertical line, and  the lognormal distributions as shaded regions. 
}
    \label{fig:lognorm_distribution}
\end{figure}

Figure~\ref{fig:lognorm_distribution} shows the column density traversed by the \Lya photons in comparison the the setup both for the lognormal setup used in \S~\ref{sec:lognormal} as well as for a homogeneous slab.
In the lognormal case, the photon-perceived distribution is broader and exhibits a tail extending to high $N_{\rm HI}$, indicating that some photons do scatter within or pass through dense regions. However, its mode shifts to lower $N_{\rm HI}$ relative to the homogeneous case, meaning that most escaping photons encountered lower column densities, simply because such regions are statistically more prevalent. This is consistent with the shift in the emergent spectral peaks observed in Fig.~\ref{fig:spectra_lognorm}, where increasing $\sigma$ leads to spectra that seem to be shaped by lower column densities than the initial distribution's mean. 

\begin{figure}
    \centering
    \includegraphics[width=0.9\linewidth]{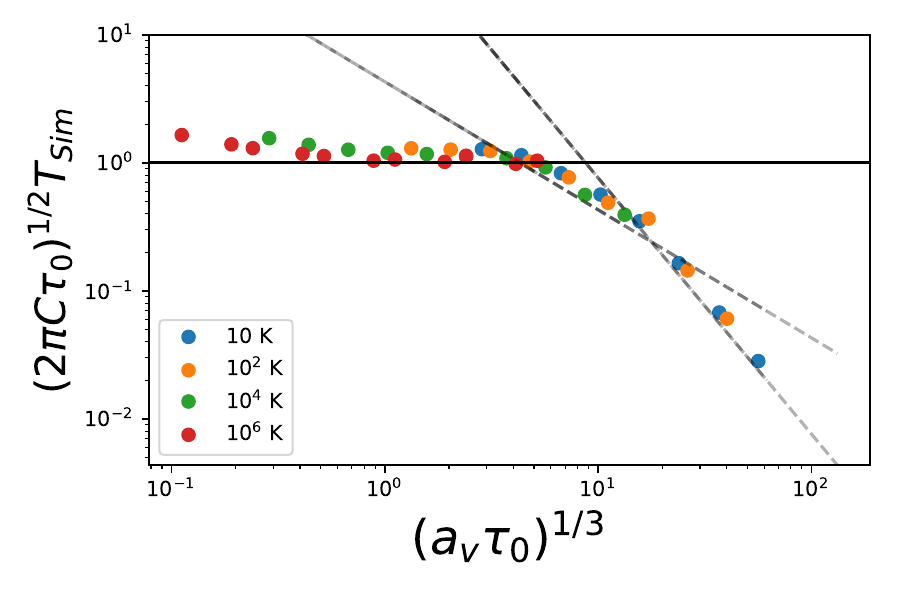}
    \caption{Transmission through an externally illuminated slab normalized to the analytical value Eq.~\ref{eq:transmission}. The additional two slopes show $(a\tau_0)^{-1/3}$ and $(a\tau_0)^{-2/3}$.}
    \label{fig:transmission}
\end{figure}

\section{Case of very large \texorpdfstring{$a_v\tau_0$}{a tau0}}
\label{sec:largeatau0}
Note that the prefactor of $\alpha \approx 1/2$ in Eq.~\eqref{eq:transmission} is an approximation. In reality photons with $x\approx x_{\rm esc}$ undergo (another) random walk with path length
\begin{equation}
    \lambda_{\rm mfp}(x=x_{\rm esc}) = \frac{L}{x_{\rm esc}} 
\end{equation}
with
\begin{equation}
    x_{\rm esc}=\left(\frac{\tau_0 a_v}{\sqrt{\pi}}\right)^{1/3}.
\end{equation}
This means one ends up with another `Gambler's ruin problem', but the Gambler leaves the casino at a fixed amount of money ($=x_{\rm esc}$). Thus, the factor of $1/2$ decreases to
\begin{equation}
    \alpha = \text{min}(1/2, 1/x_{\rm esc})
\end{equation}
which translates to $T\sim \tau_0^{-1/2}(a_v\tau_0)^{-1/3}$.
Note however, that this does not translate to a larger central flux -- as the photons are already at $x\approx x_{\rm esc}$. Thus, while the above consideration does nominally lead to a higher $\tilde f$, the consideration is somewhat only of academic interest here.

Fig.~\ref{fig:transmission} shows the transmission probability through an externally illuminated slab. One can see that while for the most part Eq.~\ref{eq:transmission} works well, for $a_v\tau_0\gtrsim 10^3$ one requires the correction discussed above. Interestingly, for even larger values of $a_v\tau_0$ the transmission drops even more. We confirmed that, also in this case, the photons escape at $x\sim x_{\rm esc}$, i.e., the flux at line center is unaffected, and a pure line-center random-walk $1/\tau_0$ solution does not seem to apply.



\bsp	
\label{lastpage}
\end{document}